\documentclass[10pt, twocolumn]{IEEEtran}

\usepackage{amsmath, amssymb, amsfonts, amssymb, amsthm}
\usepackage{bbm}
\usepackage{graphicx} \graphicspath{ {figure/} }
\usepackage{epstopdf} 
\usepackage[noabbrev]{cleveref}
\usepackage{scalefnt}
\usepackage{indentfirst}
\usepackage{cite}
\usepackage{bm}
\usepackage{enumerate}
\usepackage{enumitem}
\usepackage[caption=false,subrefformat=parens,labelformat=parens]{subfig}
\usepackage{scalerel}
\usepackage{scalerel}
\usepackage{algpseudocode}
\usepackage{multirow}
\usepackage{caption}
\usepackage[table,xcdraw]{xcolor}

\usepackage{algorithm}
\usepackage{algcompatible}
\renewcommand{\algorithmiccomment}[1]{\hfill ~#1}
\algnewcommand\INPUT{\item[\textbf{Input:}]}
\algnewcommand\INITIAL{\item[\textbf{Initialization:}]}
\algnewcommand\OUTPUT{\item[\textbf{Output:}]}
\algnewcommand\RETURN{\item[\textbf{Return:}]}
\algnewcommand\ITER{\item[\textbf{Iteration:}]}
\algrenewcommand\algorithmiccomment[2][\small]{{#1\hfill\ #2}}

\theoremstyle{plain}

\renewcommand{\algorithmiccomment}[2][.5\linewidth]{\leavevmode\hfill\makebox[#1][l]{//~#2}}

\theoremstyle{definition}

\theoremstyle{remark}

\begin{document}
\IEEEoverridecommandlockouts

\title{Towards Deep Learning-aided Wireless Channel Estimation and Channel State Information \\ Feedback for 6G}

\author{Wonjun~Kim, Yongjun~Ahn, Jinhong~Kim, and~Byonghyo~Shim\\  Seoul National University, Seoul, Korea}

\maketitle

\begin{abstract}
Deep learning (DL), a branch of artificial intelligence (AI) techniques, has shown great promise in various disciplines such as image classification and segmentation, speech recognition, language translation, among others. This remarkable success of DL has stimulated increasing interest in applying this paradigm to wireless channel estimation in recent years. Since DL principles are inductive in nature and distinct from the conventional rule-based algorithms, when one tries to use DL technique to the channel estimation, one might easily get stuck and confused by so many knobs to control and small details to be aware of. The primary purpose of this paper is to discuss key issues and possible solutions in DL-based wireless channel estimation and channel state information (CSI) feedback including the DL model selection, training data acquisition, and neural network design for 6G. Specifically, we present several case studies together with the numerical experiments to demonstrate the effectiveness of the DL-based wireless channel estimation framework.
\end{abstract}

\section{Introduction}
Artificial intelligence (AI) is a powerful tool to perform tasks that seem to be simple for human being but are extremely difficult for conventional (rule-based) computer program.
Deep learning (DL), a branch of AI techniques popularized by Lecun, Bengio, and Hinton~\cite{DL_nature}, has shown great promise in many practical applications.
In the past few years, we have witnessed great success of DL in various disciplines such as traditional Go game, image classification, speech recognition, language translation, to name just a few~\cite{Deep_residual_learning, Go_game, DL_textbook}.
DL techniques have also shown excellent prominence in various wireless communication tasks such as active user detection (AUD), spectrum sensing, and resource scheduling~\cite{hkjoo, kjsuh, yjahn_ictc}.
Also, efforts are underway to exploit DL as a means to overcome the various problems in the wireless channel estimation, particularly for the channel estimation in the high frequency bands (e.g., pilot overhead and channel estimation accuracy degradation in mmWave MIMO systems).
Notable applications include multiple-input-multiple-output (MIMO) channel estimation, angle-domain channel estimation for mmWave and terahertz (THz) band, and channel estimation in vehicular to everything (V2X) system~\cite{DLCE_example1, DLCE_example2, DLCE_example3}.
In response to this trend, 3GPP recently decided to include artificial intelligence as one of main item in 5G advanced (Rel 17)~\cite{TR_2022}.

When one tries to apply DL technique to the wireless channel estimation, one can be easily overwhelmed by so many knobs to control and small details to be aware of.
In contrast to the conventional linear minimum mean square error (LMMSE) or least square (LS) techniques where the estimation algorithm design is fairly straightforward, DL-based technique requires lots of hands-on experience and heuristic tips and tricks in the design of neural network, training dataset generation, and also choice of the training strategy.
In fact, since the DL-aided system is data-driven and inductive in nature, one might easily get confused, stuck in the middle, or come up with suboptimal channel estimation scheme.

The primary goal of this paper is to show that DL is an effective means for the channel estimation and feedback, in particular for future wireless systems using high frequency band.
Recently, there have been some studies on the AI-based channel estimation~\cite{DLCE_example1, DLCE_example2, DLCE_example3}.
Since the main focus of these studies is to present the specific DL-based channel estimation technique for a specific wireless scenario (e.g., channel estimation in the MIMO system and optical wireless system), it might not be easy to grasp general idea and systematic view on the problem.
In a nutshell, successful design of DL-based channel estimation and CSI feedback scheme comes down to 1) the choice of a proper DL model for the target wireless system, 2) detailed deep neural network architecture design, and 3) training data acquisition along with training strategy selection.
We get to the heart of these and discuss detailed issues and provide several useful design tips learned from our past experiences in the time/frequency-domain channel estimation, geometric channel parameter estimation (e.g., angle-of-departure/arrival (AoD/AoA) estimation) in mmWave systems, and channel state information (CSI) feedback.

\begin{figure*}
	\centering
	\includegraphics[width=1.7\columnwidth]{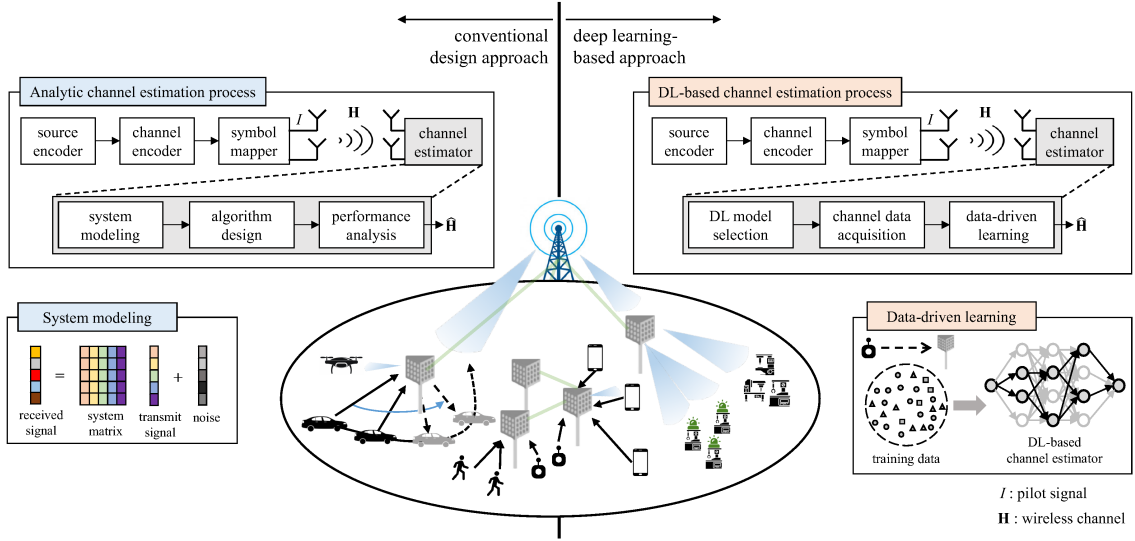}
	\caption{Design principles of traditional channel estimation algorithm and DL-based channel estimation technique.}
	\label{fig_1}
\end{figure*}

The rest of this paper is organized as follows.
In Section II, we briefly review the design principles of conventional and discuss learning techniques used to estimate and exploit the wireless channel.
In Section III, we explain the training dataset collection and neural network architecture design issues.
In Section IV, we provide several case studies that use DL to estimate and exploit channels in the wireless system, together with plenty of experimental results.
We discuss future issues and conclude the paper in Section V.

\section{Basics of Deep Learning-Based Wireless Channel Estimation}
In this section, we briefly compare two design principles: conventional channel estimation and the DL-based channel estimation techniques. 
We then discuss how DL technique can be mapped to the channel estimation in specific wireless environments.

\subsection{Design Principle of Conventional Wireless Channel Estimation}
When designing an algorithm to estimate the wireless channel, one should perform the system modeling, algorithm design, and performance analysis (see Fig.~\ref{fig_1}).
System model, typically expressed as a clean-cut linear equation, defines the relationship between observation (e.g., received pilot signal) and variables to be recovered (e.g., time-domain channel tap, angle of arrival/departure).
Using this, a proper algorithm achieving near optimal performance is developed (e.g., MMSE channel estimator, expectation-maximization (EM)-based channel estimator) and the theoretic analysis is conducted to obtain the performance bound such as the mean squared error (MSE) bound.

One well-known example is the LMMSE channel estimator derived from the linear system model given by~\cite{shpark}
\begin{align}
    \mathbf{y} = \mathbf{X}\mathbf{h} + \mathbf{n}, \label{eq:linear}    
\end{align}
where $\mathbf{y} \in \mathbb{C}^{m \times 1}$ is the received signal vector, $\mathbf{X} \in \mathbb{C}^{m \times n}$ is a matrix containing the pilots, $\mathbf{h} \in \mathbb{C}^{n \times 1}$ is the channel vector to be reconstructed, and $\mathbf{n}\sim \mathcal{CN}(\mathbf{0},\sigma_{n}^{2} \mathbf{I})$ is a complex Gaussian noise vector.
The linear estimator $\mathbf{W}^{\ast}$ that minimizes the Bayesian MSE $E\left[ \left( \mathbf{h} - \hat{\mathbf{h}} \right)^2 \right] = E\left[ \left( \mathbf{h} - \mathbf{W}^{H} \mathbf{y} \right)^2 \right]$ is
\begin{align}
    \mathbf{W}^{\ast} = \arg\min_{\mathbf{W}} \text{MSE}(\mathbf{W}).
\end{align}
After taking the gradient $\frac{\partial \text{MSE}}{\partial \mathbf{W}} = 2E\left[\mathbf{y}\mathbf{y}^{H}\right] \mathbf{W} - 2E\left[\mathbf{y}\mathbf{h}^{H}\right]$ and setting it to zero, we have
\begin{align}
    \mathbf{W}^{\ast} &= E\left[\mathbf{y}\mathbf{y}^{H}\right]^{-1} E\left[\mathbf{y} \mathbf{h}^{H} \right]  \\
    &= E\left[ \mathbf{X}\mathbf{h}\mathbf{h}^{H}\mathbf{X}^{H} + \mathbf{n}\mathbf{n}^{H}\right]^{-1} E\left[ \mathbf{X}\mathbf{h}\mathbf{h}^{H} + \mathbf{n}\mathbf{h}^{H}\right] \\
    &= \left(\mathbf{X}R_{hh} \mathbf{X}^{H} + \sigma_{n}^{2}\mathbf{I} \right)^{-1} \mathbf{X} R_{hh},
    \label{eq:W}
\end{align}
where $R_{hh} = E[\mathbf{h}\mathbf{h}^{H}]$ is the autocorrelation matrix of $\mathbf{h}$.
Let $\hat{\mathbf{h}} = (\mathbf{W}^{\ast})^{H} \mathbf{y}$, then
\begin{align}
    \hat{\mathbf{h}} &= R_{hh} \mathbf{X}^{H} \left( \mathbf{X} R_{hh} \mathbf{X}^{H} + \sigma_{n}^{2} \mathbf{I} \right)^{-1} \mathbf{y}.
\end{align}
Note that, when $\mathbf{X}$ is a diagonal matrix, we also obtain $\hat{\mathbf{h}} = R_{hh} \left( R_{hh}+\sigma_{n}^{2} (\mathbf{X}\mathbf{X}^{H})^{-1} \right)^{-1} \tilde{\mathbf{h}}$ where $\tilde{\mathbf{h}} = \mathbf{X}^{-1} \mathbf{y}$ is the LS estimate of $\mathbf{h}$. 
LMMSE estimate is a channel estimate minimizing the MSE when the system is expressed as an overdetermined linear where the number of measurements is larger than or equal to the size of unknown channel vector ($m \ge n$).
However, the task to recover the channel vector in an underdetermined scenario where the measurement size is smaller than the unknown channel vector ($m<n$) is challenging and problematic, since one cannot find out the unique solution in general.
If one wish to convert this ill-posed problem into a well-posed one, an additional side information is indispensable.

\begin{figure*}[t]
	\centering
	\subfloat[]{\includegraphics[width=1.85\columnwidth]{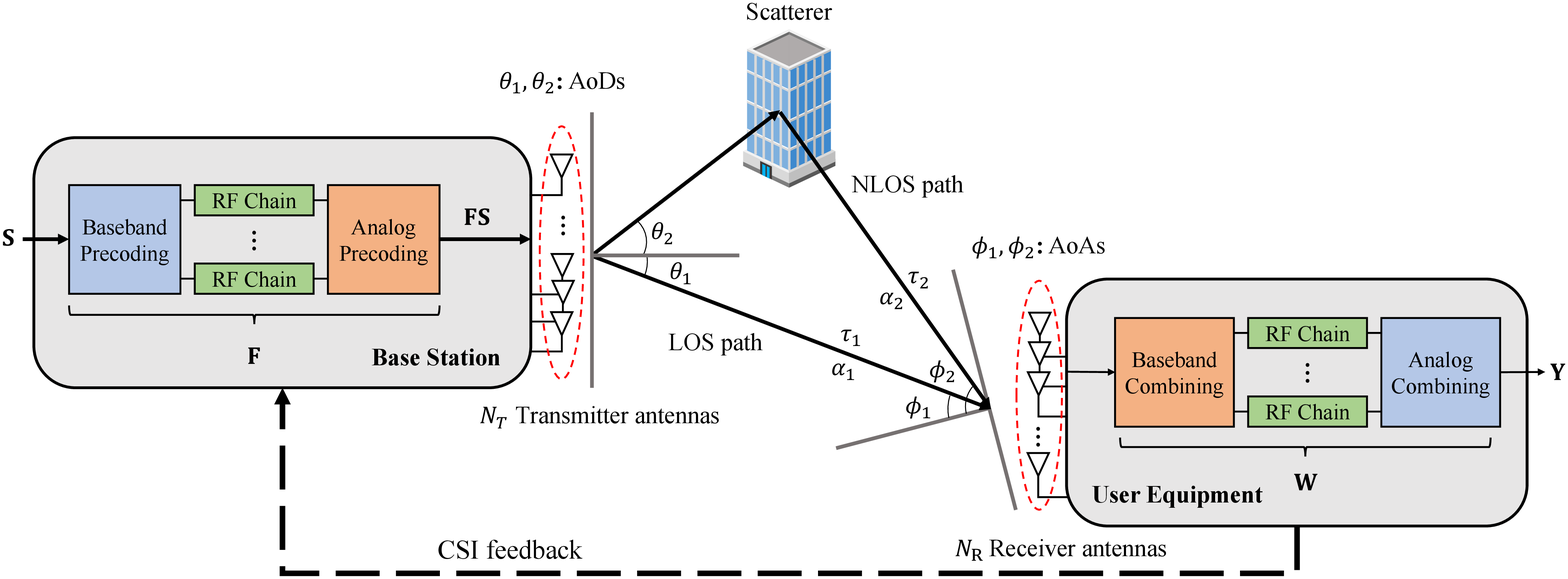}}
	\hfill
	\subfloat[]{\includegraphics[width=1.7\columnwidth]{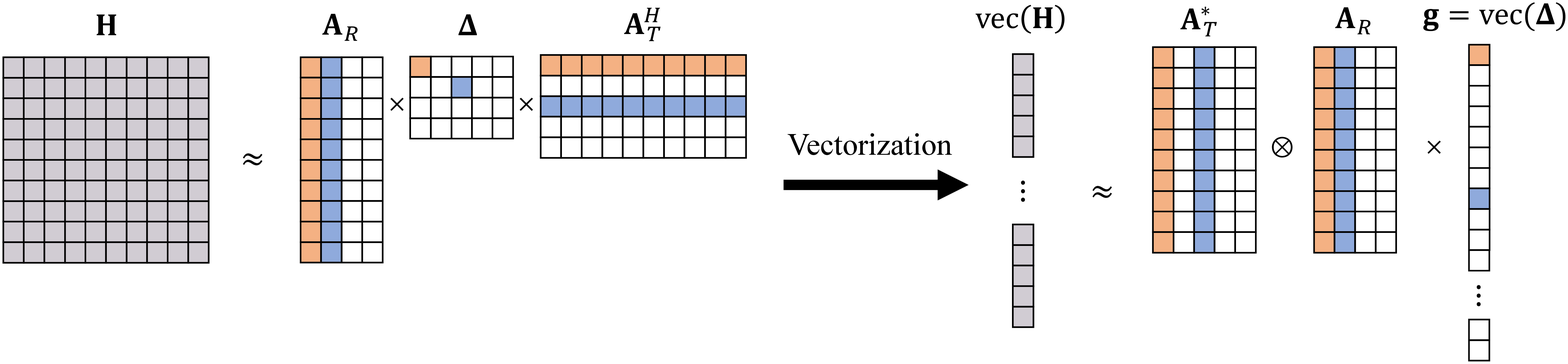}}
	\hfill
	\caption{(a) Illustration of mmWave MIMO system with $N_T$ transmission antennas and $N_R$ receiver antennas. (b) Illustration of approximated channel using discretized angular bases $\mathbf{A}_R$ and $\mathbf{A}_T$.}
	\label{fig_CS}
\end{figure*}
When the sparsity constraint is provided as a side information, compressed sensing (CS)-based technique can be used to estimate the channel in the underdetermined system.
For example, a mmWave propagation channel (e.g., FR2 in 5G~\cite{TR_FR2}) is characterized by the geometric parameters such as AoD/AoA, path delay, and path gain.
Due to the the severe attenuation of signal power caused by the high diffraction and penetration loss, atmospheric absorption, and rain attenuation in the mmWave band, the number of effective paths is at most a few (LoS and at most one or two NLoS paths), meaning that the channel can be represented by a small number of geometric parameters~\cite{jhkim1}.
Using this property as a side information, one can find out the sparse parameters used to reconstruct the mmWave channel.

Consider the narrowband geometric mmWave MIMO channel model given by
\begin{align}
\mathbf{H} = \sum^{P}_{i=1} \alpha_{i} e^{-j 2\pi \tau_i f_s }\mathbf{a}_R(\theta_{i}) \mathbf{a}_T^H(\phi_{i}),
\label{eq:MIMO_channel}
\end{align}
where $P$ is the number of propagation paths and $\alpha_{i}$, $\theta_{i}$, $\phi_{i}$, and $\tau_{i}$ are the path gain, AoA, AoD, and path delay of $i$-th path, respectively.
Also, $f_s$ is subcarrier spacing and $\mathbf{a}_{R}(\theta_{i}) \in \mathbb{C}^{N_R \times 1}$ and $\mathbf{a}_{T}(\phi_{i}) \in \mathbb{C}^{N_T \times 1}$ are the array response vectors of BS and UE, respectively ($N_T$ and $N_R$ are the number of transmit and receive antennas)\footnote{$\mathbf{a}_R(\theta_{i}) = [1, \ e^{j2\pi d\sin(\theta_{i})/\lambda}, \cdots, e^{j(N_R-1)2\pi d\sin(\theta_{i})/\lambda} ]^T$ and $\mathbf{a}_T(\phi_{i}) = [1, \ e^{j2\pi d\sin(\phi_{i})/\lambda}, \cdots, e^{j( N_T-1)2\pi d\sin(\phi_{i})/\lambda} ]^T$, respectively, where $\lambda$ is the wavelength and $d$ is the antenna spacing.}.
In the downlink transmission scenario where UE uses $i$-th combining vector $\mathbf{w}_i$ and BS uses $j$-th precoding vector $\mathbf{f}_j$, the received signal $y_{i,j}$ is
\begin{align}
y_{i,j} = \mathbf{w}_i^H \mathbf{H} \mathbf{f}_j s_j + v_{i,j},
\end{align}
where $s_j$ is transmitted pilot symbol, $\mathbf{H}\in\mathbb{C}^{N_{R}\times N_{T}}$ is the downlink channel matrix, and $v_{i,j}$ is noise.
Assuming that we use $M$ precoding vectors and $N$ combining vectors for the channel estimation, then the received signal matrix $\mathbf{Y} \in \mathbb{C}^{N \times M}$ can be expressed as
\begin{equation}
\mathbf{Y}=\mathbf{W}^{H}\mathbf{H}\mathbf{F}\mathbf{S}+\mathbf{V},
\label{eq:systemmodel}
\end{equation}
where $\mathbf{F}=[\mathbf{f}_{1},\cdots,\mathbf{f}_{M}]\in\mathbb{C}^{N_{T}\times M}$ is the beamforming matrix, $\mathbf{W}=[\mathbf{w}_{1},\cdots,\mathbf{w}_{N}]\in\mathbb{C}^{N_{R}\times N}$ is the combining matrix, $\mathbf{S}=\text{diag}(s_{1},\cdots,s_{M})$ is the matrix containing the pilot symbols, and $\mathbf{V} \in \mathbb{C}^{N \times M}$ is the matrix containing the noise elements (see Fig.~\ref{fig_CS}(a)).

In order to estimate the channel, CS-based techniques approximate $\mathbf{H}$ using discretized model $\mathbf{H} \approx \mathbf{A}_R \mathbf{\Delta} \mathbf{A}_T^H$, where $\mathbf{A}_R \in \mathbb{C}^{N_R \times G_r}$ and $\mathbf{A}_T \in \mathbb{C}^{N_T \times G_t}$ are uniformly discretized angular basis for AoA and AoD, respectively\footnote{$\mathbf{A}_R = \frac{1}{\sqrt{N_R}}[\mathbf{a}_\mathrm{R}(0), \mathbf{a}_\mathrm{R} \left( \frac{2\pi}{G_r} \right), \cdots,  \mathbf{a}_\mathrm{R} \left( \frac{2\pi}{G_r}(G_r - 1) \right)]$ and $\mathbf{A}_T = \frac{1}{\sqrt{N_T}}[\mathbf{a}_\mathrm{T}(0), \mathbf{a}_\mathrm{T} \left( \frac{2\pi}{G_t} \right), \cdots,  \mathbf{a}_\mathrm{T} \left( \frac{2\pi}{G_t}(G_t - 1) \right)]$, respectively, where $G_r$ and $G_t$ denote the number of angular grids in each basis~\cite{alkhateeb2014channel, li2017millimeter, EM_CE}. $\mathbf{\Delta}$ is a sparse path gain matrix containing $P$ non-zero channel path gains. Specifically, the $(i,j)$-th element of $\mathbf{\Delta}$ is the path gain corresponding to the $i$-th angular grid for the AoA and the $j$-th angular grid for the AoD.} (see Fig.~\ref{fig_CS}(b)).
To convert the channel estimation problem into the \textit{sparse signal recovery problem}, we rearrange the system model in~\eqref{eq:systemmodel} using the mixed Kronecker matrix-vector property\footnote{$\text{vec}(\mathbf{AXC}) = (\mathbf{C}^T \otimes \mathbf{A})\text{vec}(\mathbf{X})$}:
\begin{align}
\tilde{\mathbf{y}} = \text{vec}(\mathbf{Y}) \approx \mathbf{\Phi} \mathbf{g} + \tilde{\mathbf{v}},
\label{eq:sparse_vector}
\end{align}
where $\mathbf{\Phi} = (\mathbf{A}^H_{T}\mathbf{F}\mathbf{S})^T \otimes (\mathbf{W}^H \mathbf{A}_R ) \in \mathbb{C}^{MN \times G_r G_t}$ is the sensing matrix, $\mathbf{g} = \text{vec}(\mathbf{\Delta}) \in \mathbb{C}^{G_r G_t \times 1}$ is the sparse path gain vector, and $\tilde{\mathbf{v}}=\text{vec}(\mathbf{V})$ is the vectorized noise matrix.
In this setup, $\mathbf{H} \approx \mathbf{A}_R \mathbf{\Delta} \mathbf{A}_T^H$ can be reconstructed as long as we know the position and value of the $P$ nonzero elements in $\mathbf{g}$.

To identify nonzero elements in $\mathbf{g}$, sparse recovery algorithms such as the orthogonal matching pursuit (OMP) can be used~\cite{choi2017compressed}.
For instance, in each iteration, the OMP algorithm picks one column $\mathbf{\Phi}_{i}$ of $\mathbf{\Phi}$ which is maximally correlated to the residual vector $\mathbf{r}^{k-1}$:
\begin{align}
    i=\arg\max_{i=1,\cdots,G_r G_t} \lVert \mathbf{\Phi}_{i}^{H} \mathbf{r}^{k-1} \rVert_{2}^{2},
\end{align}
where $\hat{\Omega}^{k-1}$, $\mathbf{r}^{k-1} = \tilde{\mathbf{y}}-\mathbf{\Phi}_{\hat{\Omega}^{k-1}} \hat{\mathbf{g}}^{k-1}$, and $\hat{\mathbf{g}}^{k-1}=(\mathbf{\Phi}_{\Omega^{k-1}}^{H} \mathbf{\Phi}_{\Omega^{k-1}})^{-1} \mathbf{\Phi}_{\Omega^{k-1}}^{H} \tilde{\mathbf{y}}$ are the estimated positions of nonzero elements until $(k-1)$-th iteration, $(k-1)$-th residual vector, and $(k-1)$-th sparse path gain vector estimate, respectively. 

While the CS-based channel estimation is effective in dealing with the sparsity of the mmWave channel, it might not work well in practical scenarios where the mismatch between the true angles $\{ \theta_i,\phi_i \}$ and the quantized angles in the angular bases $\{\hat{\theta}_{i} \in [0,\frac{2\pi}{G_r},\cdots,\frac{2\pi}{G_r} (G_r-1)] ,\hat{\phi}_i \in [0,\frac{2\pi}{G_t},\cdots,\frac{2\pi}{G_t} (G_t-1)]\}$ is considerable~\cite{jhkim1}.
By applying high-resolution angle quantization, one can reduce the error caused by the mismatch.
In this case, however, the column dimension of the sensing matrix $\mathbf{\Phi}$ would be much larger than the size of measurement vector $\tilde{\mathbf{y}}$, increasing the underdetermined ratio of the system and degrading the channel estimation performance.

\subsection{Design Principle of DL-based Wireless Channel Estimation}
As discussed, conventional channel estimation techniques are effective in a system where input-output relationship is explicit and linearly expressed.
However, efficacy of this approach can be degraded when the wireless environments and systems are becoming complicated and the input-output relationship is nonlinear.

As an entirely-new paradigm to deal with the channel estimation problem, DL has been received much attention in recent years.
When a system is complex, meaning that it has complicated inputs/outputs relationship and the internal structure is highly nonlinear, it would be very difficult to come up with a closed-form analytic solution.
In fact, when the system has a constraint (e.g., solution lies on a nonlinear manifold or output is nonlinear transformation of input) and thus analytic solution is far from being optimal, DL comes to the rescue.
A holy grail of DL is to let machine learn the complicated, often highly nonlinear, relationship between the input dataset and the desired output without human intervention.
In a nutshell, DL-based channel estimation is distinct from the conventional channel estimation in two main respects: \textit{data-driven training} and \textit{end-to-end learning} of the black box (see Fig.~\ref{fig_1}).
Instead of following the analytical avenue, the DL model approximates the channel estimation function via the training process.
In the training phase, DL parameters (weights and biases) are updated to identify the end-to-end mapping between the input dataset (typically received pilot signal) and the wireless channel.
Once the training is finished, in the inference phase, DL returns the wireless channel for the input signal.
One can infer that what we essentially need to do is \textit{to feed well-prepared training dataset} into the \textit{properly designed neural network}.
It seems to be simple but requires lots of hands-on experience to achieve the maximum effect.

\subsection{Learning Techniques for DL-based Wireless Channel Estimation and Exploitation}
When one tries to use DL to estimate the wireless channel, perhaps the first thing to consider is to determine what learning technique to use.
Depending on the design goal, training dataset, and learning mechanism, DL techniques can be roughly divided into three categories: supervised learning, unsupervised learning, and reinforcement learning.

\begin{table*}[]
\centering
\caption{Summary of DL Techniques for Channel Estimation and Exploitation}
\resizebox{.99\textwidth}{!}{
\begin{tabular}{c|l|l|l}
\hline
\rowcolor[HTML]{C0C0C0} 
\textbf{Learning technique}                                                     & \multicolumn{1}{c|}{\cellcolor[HTML]{C0C0C0}\textbf{Applicable problem}}                      & \multicolumn{1}{c|}{\cellcolor[HTML]{C0C0C0}\textbf{Loss function}}                           & \multicolumn{1}{c}{\cellcolor[HTML]{C0C0C0}\textbf{Application example}}                                                              \\ \hline
                                                                                & \begin{tabular}[c]{@{}l@{}}Detection problem using\\ the classification training\end{tabular} & \begin{tabular}[c]{@{}l@{}}Cross entropy,\\ Kullback-Leibler (KL) divergence,\end{tabular}    & \begin{tabular}[c]{@{}l@{}}LoS/NLoS detection,\\ Time-domain channel tap detection\end{tabular}                                                      \\ \cline{2-4} 
\multirow{-3}{*}{\begin{tabular}[c]{@{}c@{}}Supervised\\ learning\end{tabular}} & \begin{tabular}[c]{@{}l@{}}Estimation problem using\\ the regression training\end{tabular}    & \begin{tabular}[c]{@{}l@{}}Mean squared error (MSE),\\ Mean absolute error (MAE)\end{tabular} & \begin{tabular}[c]{@{}l@{}}AoA, AoD, DoA estimation,\\ Path gain estimation\end{tabular}                                   \\ \hline
\begin{tabular}[c]{@{}c@{}}Unsupervised\\ learning\end{tabular} &
Optimization problem
& \begin{tabular}[c]{@{}l@{}}Objective function to be optimized\\ (e.g., sum-rate, cell throughput) \end{tabular}
& \begin{tabular}[c]{@{}l@{}}Pilot signal allocation,\\Channel state information feedback \end{tabular}                    \\ \hline
\begin{tabular}[c]{@{}c@{}}Reinforcement\\ learning\end{tabular}
& \begin{tabular}[c]{@{}l@{}}Sequential decision making \\ problem\end{tabular}
& \begin{tabular}[c]{@{}l@{}}Cumulative reward\\ (e.g., sum-rate, total power consumption) \end{tabular}
& \begin{tabular}[c]{@{}l@{}}Beam tracking and selection,\\ CQI feedback\end{tabular} \\ \hline
\end{tabular}
}
\end{table*}

1) \textbf{Supervised learning}:
primary goal of the supervised learning is to learn a mapping function between the input dataset and the desired solution called \textit{label}.
To scrutinize the quality of a designed neural network and reflect it in the weight update process, we need a loss function that measures how far the predicted channel (or latent variable constructing channels) is from the label.
The difference between two, in a form of MSE or cross entropy, is used as a loss function.
Typically, there are two types of the supervised learning: \textit{classification} to find out the categorical class of given input and \textit{regression} to return the numerical value.
The classification task is suitable for the detection problem such as the time-domain channel tap detection, LoS/NLoS detection, and indoor/outdoor detection and the regression task is a good fit for the estimation problem such as the angle (DoA/AoD) estimation, path gain estimation, and path delay estimation~\cite{Indooroutdoor}.

\begin{figure}
	\centering
	\includegraphics[width=.98\columnwidth]{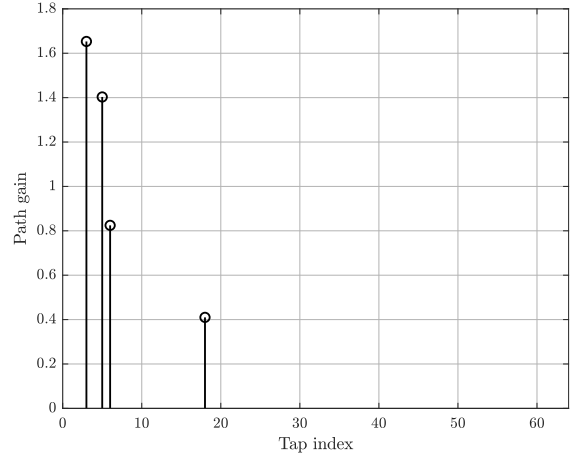}
	\caption{Illustration of tapped delay lines of the time-domain channel. In this case, we set $\Omega = [0, 0, 1, 0, 1, 1, 0, \cdots, 1, \cdots, 0]$.}
	\label{fig:tap_delay_example}
\end{figure}

In the time-domain channel tap detection, for example, we identify a few nonzero (dominating) taps among all possible tapped delay-lines so that the problem can be well modeled as a multi-label classification problem identifying a few, say $k$, labels among $N$ classes.
By employing the set of received vectors $\left\{\mathbf{y}_{1}, \cdots, \mathbf{y}_{n}\right\}$ as inputs and the nonzero tap index $\Omega$ as an output, deep neural network (DNN) is trained to find out indices of nonzero taps (see Fig.~\ref{fig:tap_delay_example}).
In the DoA estimation problem, on the other hand, desired task is to produce the real-valued angle estimate $\hat{\mathbf{\psi}}$ from the received pilot signals $\mathbf{y}$ (see Fig.~\ref{fig_CS}(a)).
Using the MSE between the real angle $\mathbf{\psi}$ and estimate $\hat{\mathbf{\psi}}$ obtained from DL as a loss function, DNN learns the regression mapping between $\mathbf{y}$ and $\mathbf{\psi}$ (see more details in Section IV.B).

2) \textbf{Unsupervised learning}:
unsupervised learning is used when the ground-truth label is unavailable for some reasons such as nonlinearity/nonconvexity of the problem.
In this case, clearly, one cannot compute the difference between the generated output and the label and thus the design goal (i.e., objective function) is used as a loss function instead.
In the pilot signal allocation problem, for example, it is very difficult to find out an optimal pilot pattern (e.g., pilot position in the resource block (RB) in 4G LTE and 5G NR) maximizing the optimizing performance metric since the problem is highly nonlinear mixed-integer programming~\cite{scheduling1}.
In this case, by employing the quality of service (QoS) function as a loss function, the DL model can be trained.
The loss function can be the target MSE (e.g., MSE between the original channel and reconstructed channel), throughput (e.g., throughput obtained by using the channel estimated from the learned pilot pattern). and so on.

3) \textbf{Reinforcement learning} (RL):
RL is a goal-oriented learning technique where an agent learns how to solve a task by trials and errors.
In the learning process, the agent observes the state of an environment, takes an action, and then receives a reward for the action.
RL is suitable for the sequential decision-making problem whose purpose is to find out a series of actions maximizing the performance metric such as data rate, energy consumption, or latency.
Recently, deep RL (DRL) has been widely used since it can effectively handle the large-scale state-action pair in dynamically varying wireless environments.
While DRL is in general not directly used for the channel estimation, it can be used for the channel related functions such as the beam tracking and channel quality indication.
For example, when one tries to track the angle in the V2X system using DRL, a base station (BS), playing the role of the agent, observes the state (e.g., distance, velocity, and angles (AoD/DoA)) and then determines the action (adjust the beam direction) maximizing the reward (e.g., minimizing the number of retransmission).

\begin{figure*}
	\centering
	\subfloat[]{\includegraphics[width=1.8\columnwidth]{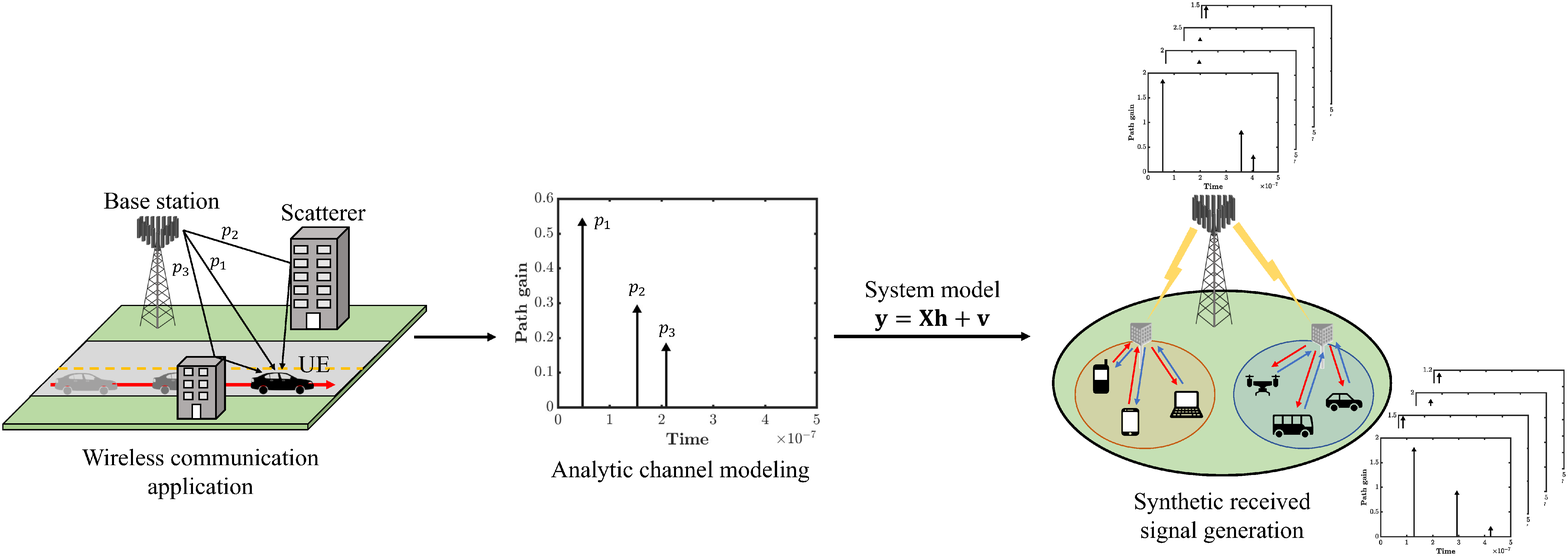}}
	\hfill
	\subfloat[]{\includegraphics[width=1.5\columnwidth]{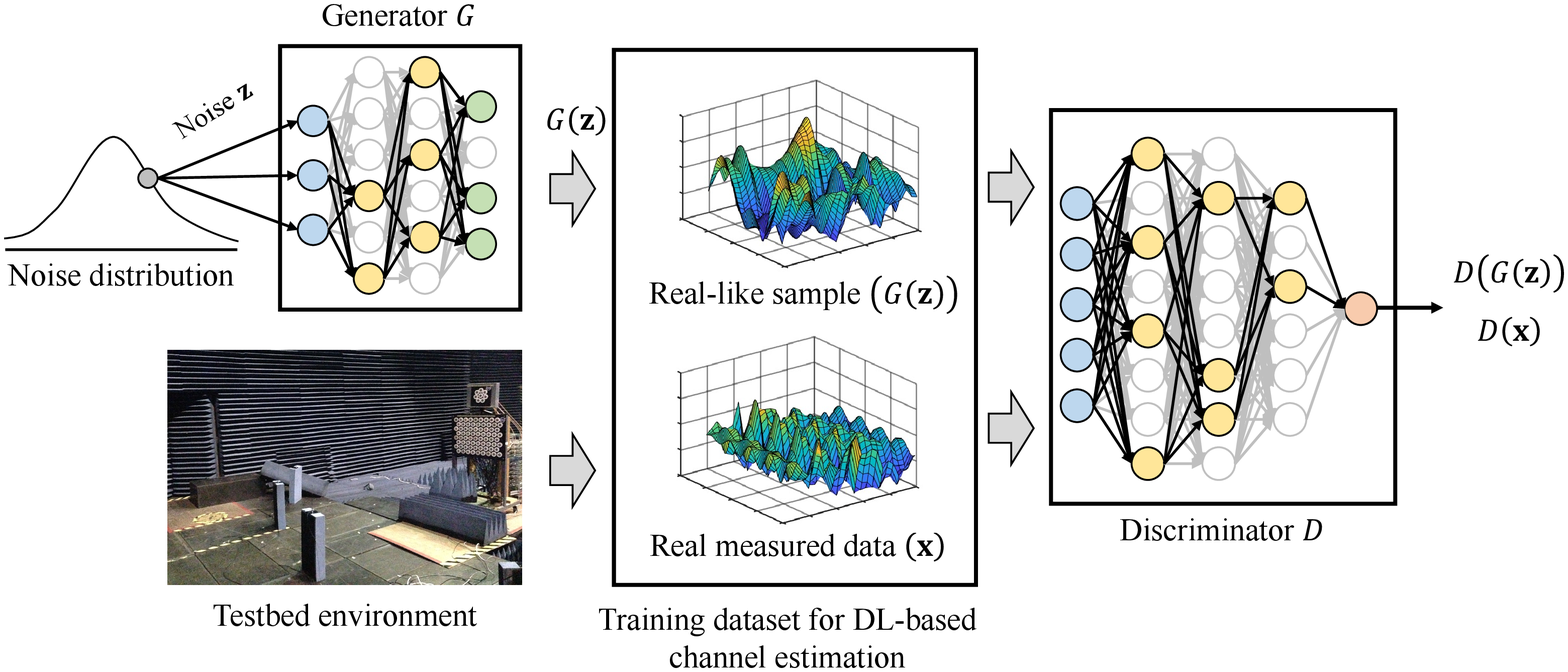}}
	\hfill
	\caption{Illustration of data acquisition strategies: (a) synthetic data generation; (b) GAN-based data generation.}
	\label{fig_2}
\end{figure*}

\section{Design Issues in DL-based Wireless Channel Estimation}
In this section, we delve into two main issues in the DL-based wireless channel estimation: sufficient and comprehensive training dataset and properly designed neural network.

\subsection{Training Dataset Acquisition}
When the number of training samples is not sufficient enough, designed DL-based channel estimation model would be closely fitted to the training dataset, making it difficult to make a reasonable inference for the unseen channel.
This problem that the trained DL model lacks the generalizing capability is often called \textit{overfitting}.
In the angle estimation task, for example, if the received signals are generated from the angles in $[0, \pi/2)$, then the trained network cannot accurately identify the rest angles in $[\pi/2, \pi)$.
To prevent the overfitting problem, a dataset should be large enough to cover all possible scenarios.
This is not easy, in particular for wireless channel, since the number of real transmissions will be humongous.
In acquiring the training dataset, we basically have three options:
\begin{itemize}
    \item Collection from the actual received signals
    \item Synthetic data generation using the analytic system model or the ray-tracing simulator
    \item Real-like training set generation using generative adversarial network (GAN)
\end{itemize}

In the training set acquisition, a straightforward option is to collect the real transmit/receive signal pair.
Doing so, however, will cause a significant overhead since it requires too many training data transmissions.
For example, when collecting one million received signals in 5G NR systems, it will take more than 15 minutes ($10^{6}$ symbols $\times$ 0.1 subframe/symbol $\times$ 8 ms/subframe).

To reduce the overhead, one can consider synthetically generated dataset (see Fig.~\ref{fig_2}(a)).
In fact, in the design, test, and performance evaluation phase of the most channel estimation algorithms, analytic models have been widely used.
For example, propagation channels such as the extended pedestrian A (EPA) channel or extended vehicular A (EVA) channel have been popularly employed in the generation of training dataset~\cite{EPA}.
In EVA channel, for example, channel is expressed by 9 tapped delays (i.e., [0, 30, 150, 310, 370, 710, 1090, 1730, 2510] ns), each of which has a different path gain in the delay-Doppler domain.
Since the synthetic data can be generated with a simple programming, time and effort to collect huge training data can be saved.
However, there might be some, arguably non-trivial, performance degradation caused by the model mismatch.

Yet another intriguing option is to use an artificial but realistic samples generated by the special DL technique.
This approach is in particular useful when the analytic model is unknown or non-existent (e.g., deep underwater acoustic channel and non-terrestrial (NTN) channel) and real measured data is not large enough (e.g., THz-band channel).
In such case, GAN technique, an approach to generate samples having the same distribution with the input dataset, can be employed~\cite{GAN}.
Basically, GAN consists of two neural networks: generator $G$ and discriminator $D$ (see Fig.~\ref{fig_2}(b)).
The generator $G$ tries to produce the real-like data samples and the discriminator $D$ tries to distinguish real (authentic) and fake data samples.
To be specific, $G$ is trained to generate real-like data $G(\mathbf{z})$ from the random noise $\mathbf{z}$ and $D$ is trained to distinguish whether the generator output $G(\mathbf{z})$ is real or fake.
In order to accomplish the mission, the min-max loss function, expressed as the cross-entropy (i.e., $H(q) = - q \log(p(q)) - (1-q) \log(1- p(q))$) between the distribution of generator output $G(\mathbf{z})$ and that of the measured channel data $\mathbf{x}$, is used:
\begin{align}
     \underset{G}{\min}\,\underset{D}{\max} \, \mathbb{E}_{\mathbf{x}}[\log(D(\mathbf{x}))] +\mathbb{E}_{\mathbf{z}}[\log(1-D(G(\mathbf{z})))],
\end{align}
where the discriminator output $D(\mathbf{x})$ is the probability of $\mathbf{x}$ being real (non-fake).
In the training process, parameters of $G$ are updated while those of $D$ are fixed and vice versa.
When the training is finished properly, the generator output $G(\mathbf{z})$ is fairly reliable so that the discriminator cannot judge whether the generator output is real or fake (i.e., $D(G(\mathbf{z})) \approx 0.5$).
This means that we can safely use the generator output for the training purpose (we will say more on GAN-based data generation in Section IV.D).

\subsection{DNN Architecture for Wireless Channel Estimation}
In the design of DNN-based channel estimator, one should consider the channel characteristics (e.g., temporal/spatial/geometric correlation), wireless environments (e.g., mmWave/THz/V2X/
UAV link), and system configurations (e.g., bandwidth, power of transmit pilot signal, number of antennas).

1) \textbf{Baseline network}:
a natural first step of the DNN design is to choose the baseline architecture.
Based on the connection shape between neighboring layers, neural network can be roughly divided into three types: fully-connected network (FCN), convolutional neural network (CNN), and recurrent neural network (RNN).

FCN can be used universally since each hidden unit (neuron) is connected to all neurons in the next layer.
When the input dataset has a spatial structure (e.g., doubly selective channel and the spatial-frequency domain channel in MIMO systems), CNN might be an appealing option.
In CNN, each neuron is computed by the convolution of the 2D spatial filter and a part (e.g., rectangular shaped region) of neurons in the previous layer.
Due to the local connectivity of the convolution filter, CNN facilitates the extraction of spatial correlated feature.
For example, in the MIMO channel estimation, spatial correlation among the uniform rectangular array (URA) antennas can be extracted using CNN.
When the input sequence is temporally correlated, which is true for most of wireless fading channels, RNN model (e.g., long-short term memory (LSTM)) might be a good choice.
In these approaches, by employing current inputs together with outputs of the previous hidden layer, temporally correlated feature can be extracted.
For instance, by applying RNN to the mmWave channel estimation, change of the Doppler frequency caused by the mobile's movement can be extracted.

Another promising network architecture distinct from the representative structures we just mentioned is the attention module~\cite{Attention}.
In essence, attention module is a network component that quantifies the correlation between every two input elements.
By measuring the attention score (a.k.a. value) between two input elements called key and query, attention module allows to model the dependencies between input elements without regard to their distance in pixel or time.
Recently, attention module is used instead of CNN and RNN since it offers the capability to extract long-distance and long-term dependencies between input elements~\cite{Attention1, Attention2}.
For example, attention module can be used to extract the spatial correlation between the distant MIMO antenna arrays (e.g., elements placed at both ends) or to extract the temporal correlation between the traffic pattern and the co-channel interference in the MIMO channel.

2) \textbf{Activation layer}:
activation layer is used to 1) embed the nonlinearity in the hidden layer and 2) generate the desired type of output in the final layer.

In each hidden layer, weighted sum of inputs passes through the activation layer to determine whether the information generated by the hidden unit is activated (delivered to the next layer) or not.
To this end, rectified linear unit (ReLU) function $f(x)=\max(x,0)$ or hyperbolic tangent function $f(x)=\tanh(x)$ can be used~\cite{DL_nature}.
By imposing the activation to the linearly transformed input, one can better promote the nonlinear operation (e.g., spline interpolation between adjacent subcarrier channel) and systematic nonlinearity (e.g., relationship between spatial-domain channel and AoA/AoD).

In the final layer of DNN, the activation layer is used to make sure that the generated output is the desired type.
In the classification problem, the ground-truth label for each class is the probability so that the final output should also be a form of the probability.
When there are several time-domain channel taps or the angular-domain channel paths are non-unique in the channel estimation problem, it would be desirable to use the sigmoid function $f(x)=\frac{1}{1+e^{-x}}$ returning the individual probability for each class~\cite{DL_nature, yjahn}.
Whereas, when the problem is modeled as a multi-class classification problem such as the CQI detection problem that generates the proper value among quantized CQI levels (e.g., 0$\sim$15 in LTE), a softmax function $f_{i}(\mathbf{x})=\frac{e^{x_{i}}}{\sum_j e^{x_{j}}}$ would be a good choice since it normalizes the output vector into the probability distribution over all classes~\cite{AUD1}.

3) \textbf{Input normalization}:
in the training process, the neural network computes the gradient of a loss function with respect to any weight and then updates the weight in the negative direction of the gradient.
When an input varies in a wide range (e.g., multi-user communication scenario), therefore, variation in the weight update process will also be large, degrading the training stability and convergence speed severely.
To prevent such ill-behavior, one should perform the normalization of the outputs for each layer.
Typically, there are two types of the normalization strategies: layer normalization and batch normalization.

First, when the input vector contains signals from multiple users with different wireless geometries, variation of the received signals would be quite large.
In such case, the layer normalization, normalizing each individual input vector, is a good choice because the layer normalization scheme ensures that the normalized input distribution has the fixed mean and variance.

Whereas, when the input data consists of several different types of information, the batch normalization (BN) might be a useful option.
In the mini-batch $\mathbf{B}=[\mathbf{y}^{(1)} \cdots \mathbf{y}^{(N)}]^{T}$ consisting of multiple input samples $\mathbf{y}^{(1)},\cdots,\mathbf{y}^{(N)}$, elements in each row of $\mathbf{B}$ (i.e., elements with the same input type) are normalized.
For example, in the DRL-based beam tracking problem, both the velocity of the moving target and the angles (AoD/DoA) are used as inputs to DRL.
Since the scale of two components would be quite different, the layer normalization will simply mess up the input dataset.
To avoid the hassle, the velocity and angles need to be normalized separately using BN.

4) \textbf{Dropout layer}:
when we use DNN consisting of multiple hidden layers, the final output is determined by the activated hidden units in each layer.
So, for the highly correlated inputs (e.g., samples generated from correlated (low-rank) MIMO channel), their activation patterns will also be similar so that the final inference can be easily corrupted in the presence of perturbations (e.g., noise, ADC quantization error, and imperfect RF filtering).
In order to mitigate this problem, the dropout layer where the activated hidden units are dropped out randomly can be used in the training phase~\cite{Dropout}.
In this scheme, by temporarily removing part of incoming and outgoing connections randomly, ambiguity (similarity) of the activation patterns among correlated dataset can be resolved.

5) \textbf{Ensemble learning}:
ensemble learning, a method to average out multiple outputs (inferences) of independently-trained networks, is conceptually analogous to the receiver diversity technique since it can enhance the output quality without requiring additional wireless resources (e.g., frequency, time, and transmitter power).
In a multi-user communication scenario, for example, the trained network might be closely fitted to the certain wireless setup (i.e., specific co-scheduled user pair) so that the trained DNN might not generate a reliable channel prediction for the inputs obtained from unobserved wireless environment.
In such case, ensemble learning becomes a useful option.
Essence of the ensemble learning is to train the multiple DNNs with different initial parameters and training sets obtained from different wireless environments and then combine the outputs of multiple DNNs.
Using the ensemble learning, one can mitigate the overfitting to the specific wireless environments and generate more robust channel estimation.

\begin{figure*}
	\centering
	\includegraphics[width=1.6\columnwidth]{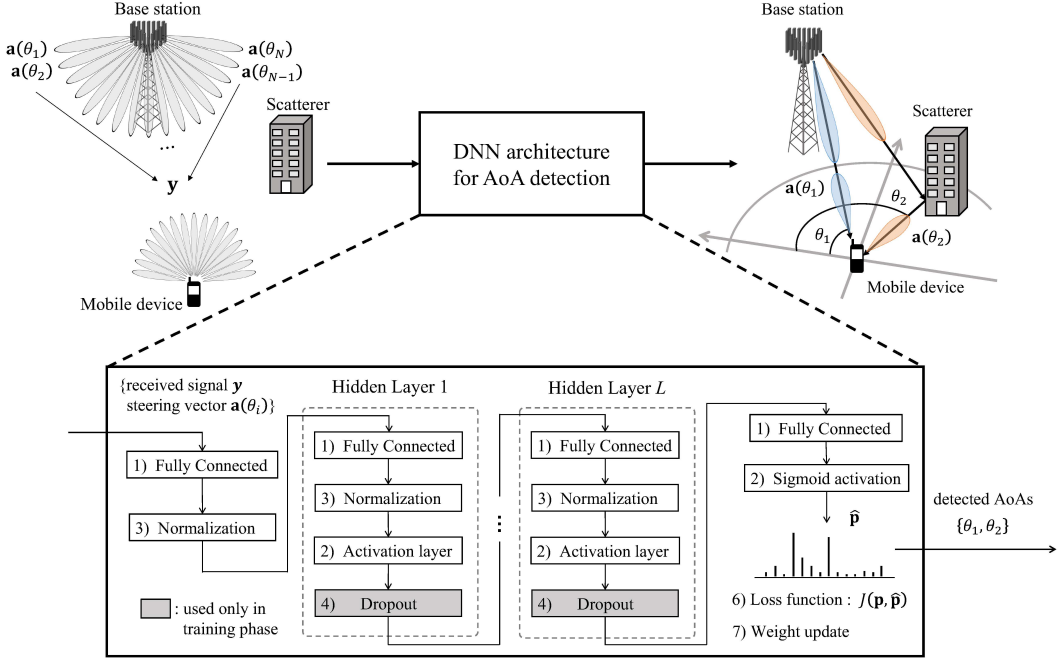}
	\caption{Exemplary DNN architecture designed for the AoA detection.}
	\label{fig_4}
\end{figure*}

6) \textbf{Loss function}:
since DNN weights are updated in a direction to minimize the loss function, the loss function should well reflect the design goal.
When the ground-truth label is available, one can use the cross entropy, MSE, or mean absolute error (MAE).
If this is not the case, as we mentioned in the unsupervised learning, one might use the design goal (e.g., throughput or bit error rate) as a loss function.

If there exist multiple goals for the problem at hand, those can be combined together.
For example, in the joint AUD and channel estimation in NOMA system, the DNN is trained to detect the active devices and at the same time estimate the channels associated with active users.
To do so, we set the loss function $J$ as the weighted sum of cross-entropy loss $J_{AUD}$ for AUD and the MSE loss $J_{CE}$ for CE (i.e., $J=J_{AUD} + \lambda J_{CE}$). 
7) \textbf{Weight update strategy}:
in order to update the network parameter set, the gradient of a loss function should be computed first.
A naive way to update the parameters is the batch gradient descent (BGD) method where the gradient of the loss function is computed for the entire training dataset.
Since the whole dataset is used in each and every training iteration, training cost is quite expensive and the training speed will be very slow.
In the non-static scenario where the channel characteristics are varying, parameters corresponding to the dynamically changing wireless environments (e.g., Doppler spread, scatter location) would not be updated properly.
A better option to deal with the issue is the stochastic gradient descent (SGD) method.
In contrast to BGD, SGD uses a small number, say $D$, of samples in each training iteration (i.e., $\Theta_{t}=\Theta_{t-1}-\frac{\eta}{D}\sum_{d=1}^{D}\nabla_{\Theta}J^{(d)}(\Theta)$) so that it can update the network parameters as soon as a few samples are obtained.

8) \textbf{Knowledge distillation}:
when one tries to use the DL-based channel estimation in the Internet of Things (IoT) device, on-device energy consumption is a big concern since most of the IoT devices are battery-powered.
To reduce the training overhead, knowledge distillation (KD), an approach to generate a relatively small-sized DL model from a trained large model, can be employed~\cite{KD}.
Key idea of KD is to train a small network (a.k.a student network) using the output of a large network (a.k.a teacher network).
In the generation of the loss function, output of the student network is compared against the output of the teacher network as well as the ground-truth label.
In doing so, the student network implemented in IoT device can easily capture the underlying feature (e.g., similarity and difference among the classes) extracted by the teacher network with minimal training overhead.

In order to exemplify the detailed techniques we discussed, we present the DNN architecture for the AoA detection (see Fig.~\ref{fig_4}).
Due to the sparse scattering in the mmWave band, a propagation path can be characterized by a few AoAs.
By identifying these angles, the receiver can align the beam direction to the transmitter, thereby maximizing the beamforming gain.
In DNN, we use the received signal $\mathbf{y}$ and the steering matrix $\mathbf{A}=[\mathbf{a}(\theta_1) \ \cdots \ \mathbf{a}(\theta_N)]$ as inputs and the set of the detected AoAs $\Omega$ as outputs\footnote{$\mathbf{a}(\theta_i)=[1 \ e^{j\pi \sin\theta_i} \ \cdots \ e^{j\pi(m-1) \sin\theta_i}]^{T}$ is the steering vector corresponding to $\theta_{i}$.}.
Since an input is a composite of the received signal and steering vectors, we use BN to normalize each component separately.
Also, to generate the individual probability for each angle, we use the sigmoid activation function in the final layer.

\begin{figure}
	\centering
	\includegraphics[width=.98\columnwidth]{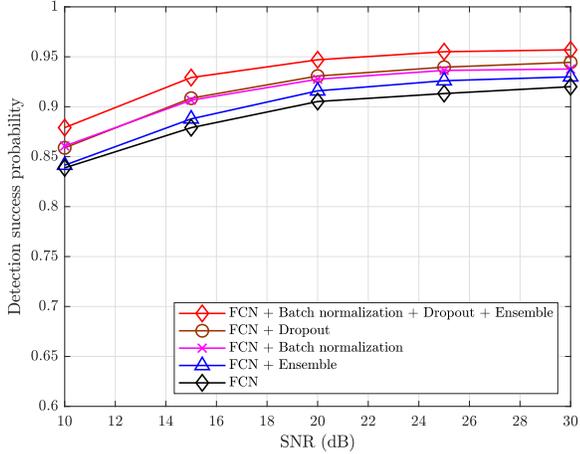}
	\caption{AoA detection performance of various DNNs as a function of SNR. The detection success probability corresponds to the percentage of detected AoAs among all angles.}
	\label{fig_5}
\end{figure}

To judge the effectiveness of the DNN components consisting of the normalization, dropout, and ensemble learning, we evaluate the success probability of the AoA detection.
In the numerical evaluations, we test 1) FCN, 2) FCN with BN, 3) FCN with the dropout layer, 4) FCN with the ensemble network, and 5) FCN with all components we discussed in this section.
As shown in Fig.~\ref{fig_5}, the performance gain introduced by the aforementioned techniques is considerable.
For example, FCN with the dropout layer achieves 4.8 dB gain over the conventional FCN since the highly correlated steering vectors can be better resolved using the dropout technique.
The gain obtained from BN is also significant (4.7 dB gain) since it reduces the variation of the received signal caused by the device location change.
Finally, when the gains induced by all techniques are combined together, we can achieve very reliable AoA detection performance, which cannot be achievable by the basic FCN even in high SNR regime.

\section{DL-aided Channel Estimation and Exploitation}
In this section, we discuss the DL-based channel estimation techniques.
We also investigate the DL-based CSI feedback in MIMO system and the meta learning-based GAN technique for real-time training. 

\subsection{Time-domain Channel Estimation for Wideband Multipath Communications}
It has been shown from recent measurement campaigns that the scattering effect of the environment is limited in many wireless channels.
In other words, propagation paths tend to be clustered within a small spread so that there exists only a few dominant scattering clusters.
In the wideband communication systems where the maximal delay and Doppler spreads are large and there are only few dominant paths, the channel can be readily modeled by a few tapped delay lines in the time-domain (i.e., delay-Doppler domain)~\cite{choi2017compressed}.
Moreover, since the path delays vary much slower than the path gains due to the temporal channel correlation, such sparsity is almost unchanged during the coherence time.
Therefore, one can greatly reduce the pilot overhead by estimating the channel in time-domain.

Let us consider the OFDM systems where we allocate the pilot symbols $\mathbf{p}_f=[p_0 \ p_1 \cdots p_{m-1}]$ in the frequency domain.
When we model the time-domain channel as a $m$-tapped channel impulse response $\mathbf{h}=\left[ h_{0} \ h_{1} \cdots h_{n-1}\right]^{T} \in \mathbb{C}^{n \times 1}$, then the received signal $\mathbf{y}$ in the frequency domain is expressed as
\begin{align}
    \mathbf{y} &= \text{diag}(\mathbf{p}_{f})\mathbf{\Phi} \mathbf{D} \mathbf{h} + \mathbf{v} \\
    &= \mathbf{P}\mathbf{h} + \mathbf{v},
\end{align}
where $\mathbf{D} \in \mathbb{C}^{n \times n}$ is the DFT matrix which plays a role to convert the time-domain channel response $\mathbf{h}$ to frequency domain response, $\mathbf{\Phi} \in \mathbb{C}^{m \times n}$ is the row selection matrix determining the location of pilots in frequency domain, and $\mathbf{P}=\text{diag}(\mathbf{p}_{f})\mathbf{\Phi} \mathbf{D}$.
Due to the limited number of scattering clusters, the number of nonzero taps $m$ in $\mathbf{h}$ is very small (i.e., $m<n$).
When we introduce the nonzero tap indicator $\boldsymbol{\delta}=[\delta_0 \ \delta_1 \cdots \delta_{n-1}]$ ($\delta_i=1$ and 0 indicate that $i$-th element of $\mathbf{h}$ is nonzero and zero, respectively), $\mathbf{y}$ can be rewritten as
\begin{align}
    \mathbf{y} = \sum_{i=0}^{n-1} \delta_i \mathbf{p}_i h_i +\mathbf{v},
\end{align}
where $h_{i}$ is the $i$-th element of $\mathbf{h}$ and $\mathbf{p}_{i}$ is the $i$-th column of $\mathbf{P}$.
Consequently, the time-domain channel estimation problem can be decomposed into two sub-problems: 1) support identification to find out nonzero positions in $\mathbf{h}$ and 2) nonzero element estimation to find out $h_i$.
Once the support of $\mathbf{h}$ is identified, estimate of the nonzero element can be easily obtained by conventional linear estimation techniques (see Section II.A).

\begin{figure*}
	\centering
	\subfloat[]{\includegraphics[width=1.17\columnwidth]{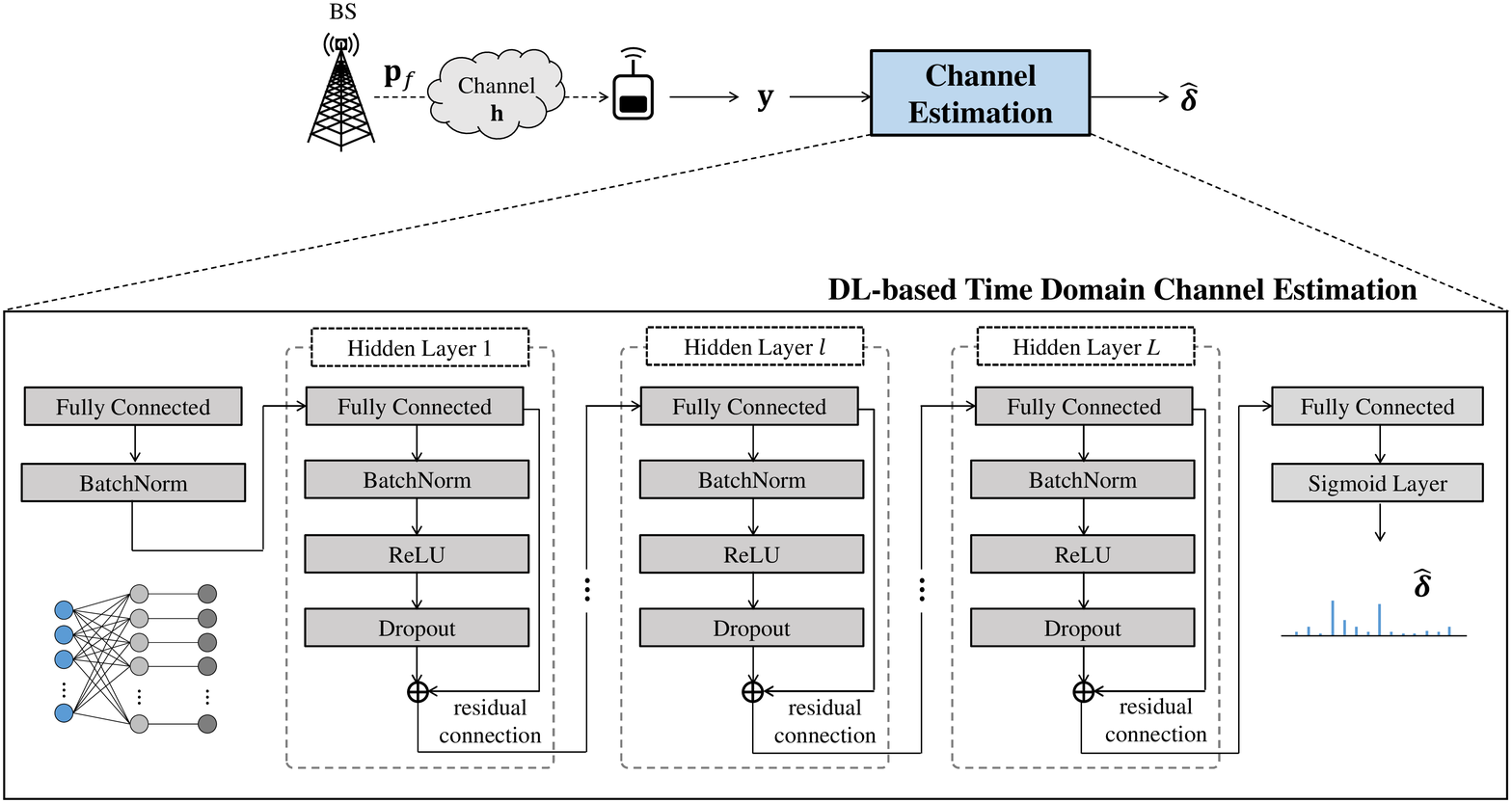}}
	\hfill
	\subfloat[]{\includegraphics[width=.8\columnwidth]{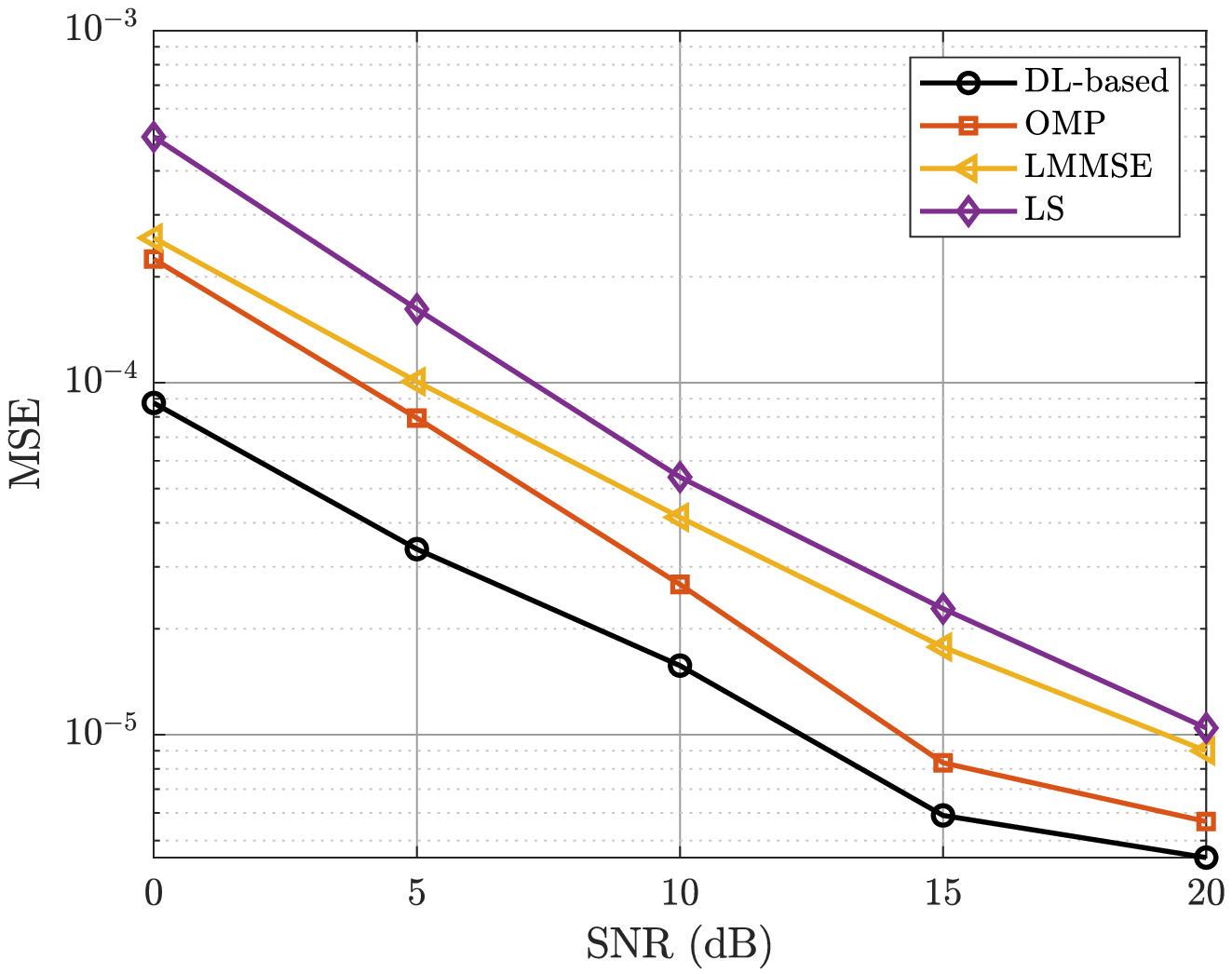}}
	\hfill
	\caption{(a) DNN architecture for time-domain channel tap detection. (b) MSE as a function of SNR ($m=64, n=256$).}
    \label{fig_6}
\end{figure*}

As mentioned, the time-domain channel tap can be detected using the CS technique, but it might not work well when the columns of the system matrix $\mathbf{P}$ are highly correlated and the sparsity (number of nonzero taps) of $\mathbf{h}$ increases.
When we use the DL technique, we need to find out the direct mapping $\boldsymbol{\delta} = g(\mathbf{y};\boldsymbol{\theta})$ between the received signal $\mathbf{y}$ and nonzero tap indicator $\boldsymbol{\delta}$, where $\boldsymbol{\theta}$ is the network parameters (see Fig.~\ref{fig_6}(a)).
In the DL-based channel tap detection, the final output is the $n$-dimensional vector $\hat{\boldsymbol{\delta}}$ whose element represents the probability of being the nonzero tap.
Thus, $\hat{\boldsymbol{\delta}}$ needs to be compared against the true probability $\boldsymbol{\delta}$ in the calculation of the loss $J$.
To do so, we employ the cross-entropy loss $J=-\sum_{i=0}^{n-1} (\delta_i \text{log}\hat{\delta_i} + (1-\delta_i)\text{log}(1-\hat{\delta_i}))$ during the training.

In Fig.~\ref{fig_6}(b), we numerically evaluate the MSE performance of the DL-based scheme as a function of SNR.
For comparison, we also examine the performance of the conventional LS estimator, LMMSE estimator, and OMP algorithm.
We observe that DL-based time-domain channel estimation outperforms the conventional schemes for all SNR regime\footnote{Note that we used 100,000 samples for the training and 5,000 samples for the test.}.
Since the DNN learns the correlation feature of the system matrix $\mathbf{P}$, DL-based channel tap detection can better discriminate the correlated supports in the test phase.
For example, we observe that the DL-based time-domain channel estimation achieves 2 dB gain over the OMP at MSE$=10^{-5}$.

\subsection{Parametric Channel Estimation for mmWave/THz MIMO Communications}
As a core technology for 5G and 6G, mmWave and THz communication have received much attention recently~\cite{rangan2014millimeter, ji2019compressed, jjpark}.
By leveraging the abundant frequency spectrum resource in mmWave/THz frequency band ($30\,\text{GHz}\sim 10\,\text{THz}$), mmWave/THz-based systems can support a way higher data rate than the conventional sub-6GHz systems. 
In the mmWave/THz systems, the beamforming technique realized by the MIMO antenna arrays is required to compensate for the severe path loss in the high frequency bands~\cite{alkhateeb2014channel}.
Since the beamforming gain is maximized only when the beams are properly aligned with the signal propagation paths, acquisition of accurate downlink channel is of importance for the success of the mmWave/THz MIMO systems.

We consider the downlink MIMO systems where the BS equipped with $N_{T}$ transmit antennas serves the user equipment (UE) equipped with $N_{R}$ receive antennas.
Then, the goal of the channel estimation problem is to obtain the downlink channel matrix of the $k$-th pilot subcarrier $\mathbf{H}[k]\in\mathbb{C}^{N_{R}\times N_{T}}$ from the received signal $\mathbf{Y}[k] \in \mathbb{R}^{M \times T}$:
\begin{equation}
\mathbf{Y}[k]=\mathbf{W}^{H}\mathbf{H}[k]\mathbf{F}\mathbf{S}[k]+\mathbf{N}[k],\quad \forall k\in\mathcal{K},
\end{equation}
where $\mathbf{F}=[\mathbf{f}_{1},\cdots,\mathbf{f}_{M}]\in\mathbb{C}^{N_{T}\times M}$ is the beamforming matrix, $\mathbf{W}=[\mathbf{w}_{1},\cdots,\mathbf{w}_{T}]\in\mathbb{C}^{N_{R}\times T}$ is the combining matrix, and $\mathbf{S}[k]=\text{diag}(s_{1}[k],\cdots,s_{M}[k])$ is the pilot symbol matrix.
While LS and LMMSE channel estimators have been popularly used for the acquisition of mmWave MIMO channel~\cite{venugopal2017channel, alkhateeb2014channel}, these conventional approaches have drawback since the amount of pilot resources is proportional to the number of transmit/receive antennas.
In fact, due to the massive numbers of transmit and receive antennas, this issue is serious for the mmWave/THz MIMO systems.
For example, if the numbers of transmit and receive antennas are $N_{T}=64$ and $N_{R}=4$, respectively, then the BS has to use at least 3 resource blocks (RBs) ($12\times 7$ resources in each RB of 4G LTE) just for the pilot transmission, occupying almost $25\%$ of a subframe in LTE.

In order to reduce the dimension of a channel matrix to be estimated, one can consider the estimation of the sparse channel parameters, not the full-dimensional MIMO channel matrix.
Specifically, when we assume the geometric narrowband channel model, then the frequency domain channel matrix $\mathbf{H}^l[k]$ of $k$-th pilot subcarrier at $l$-th channel coherence interval is expressed as
\begin{align}
\mathbf{H}^l[k] = \sum^{P}_{i=1} \alpha_{i}^l e^{-j 2\pi \tau_i^l kf_s }\mathbf{a}_R(\theta_{i}^l) \mathbf{a}_T^H(\phi_{i}^l),
\label{eq:MIMO_channel}
\end{align}
where $f_s$ is subcarrier spacing, $P$ is the number of propagation paths, $\alpha_{i}^l$ is the complex gain, $\theta_{i}^l$ and $\phi_{i}^l$ are AoA and AoD of the $i$-th path, respectively, $\tau_{i}^l$ is the path delay of $i$-th path at $l$-th channel coherence interval.

\begin{figure*}
	\centering
	\subfloat[]{\includegraphics[width=1.15\columnwidth]{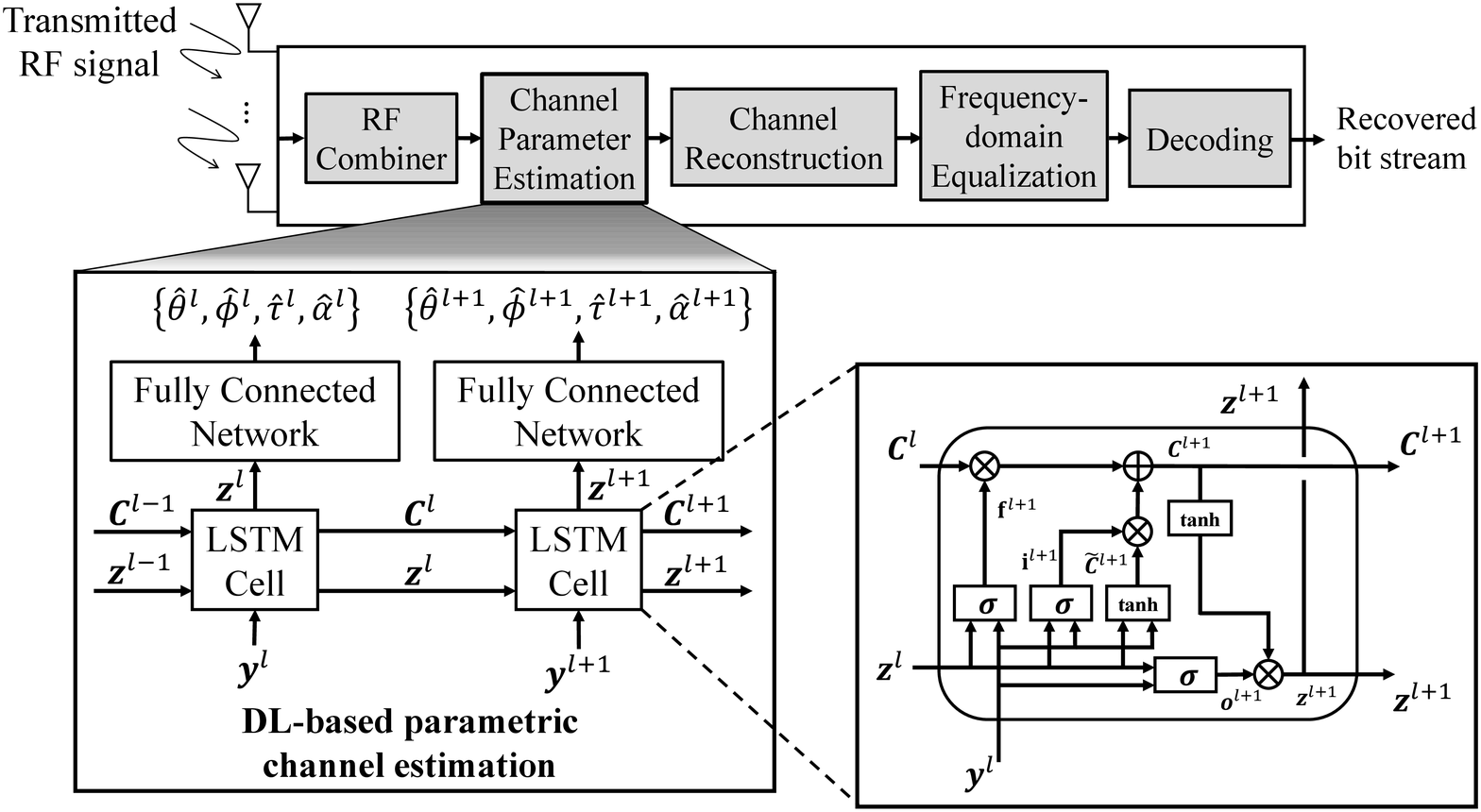}}
	\hfill
	\subfloat[]{\includegraphics[width=.82\columnwidth]{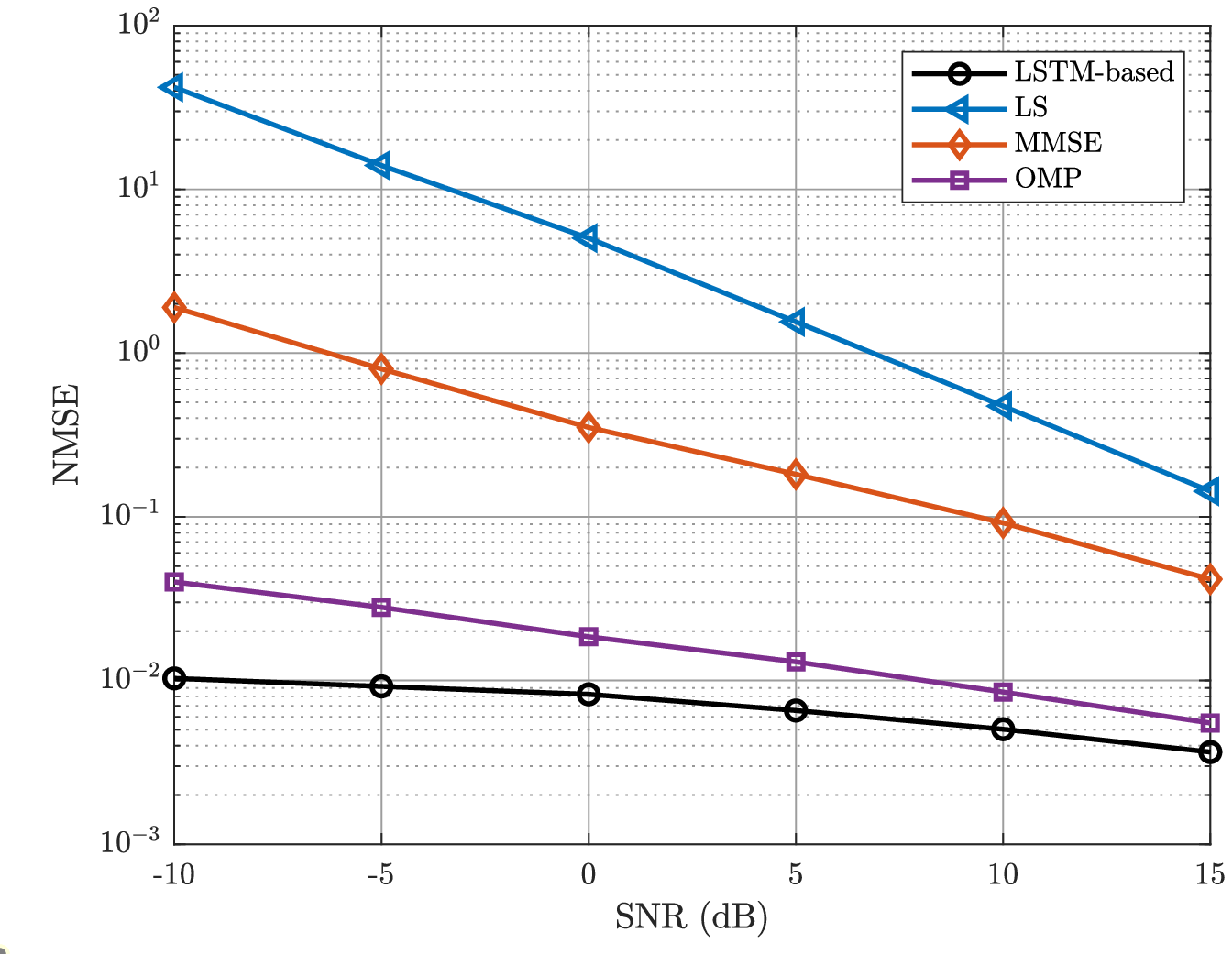}}
	\hfill
	\caption{(a) Overall receiver structure of the LSTM-based channel estimation scheme. (b) NMSE performance of channel estimation techniques as a function of SNR.}
	\label{fig_7}
\end{figure*}

One can see that $\mathbf{H}^{l}[k]$ is parameterized by a few channel parameters: AoAs $\lbrace \theta_{i}^{l}\rbrace_{i=1}^{P}$, AoDs $\lbrace \phi_{i}^{l}\rbrace_{i=1}^{P}$, path delays $\lbrace \tau_{i}^{l}\rbrace_{i=1}^{P}$, and path gains $\lbrace \alpha_{i}^{l}\rbrace_{i=1}^{P}$ since the number of effective paths is at most a few (LoS and $1\sim 2$ NLoS paths) in the mmWave/THz channel.
Typically,  the number of propagation paths $P$ is much smaller than the total number of antennas $N=N_{T}\times N_{R}$ (e.g., $P=2\sim 8$ while $N=16\sim 256$) due to the high path loss and directivity of mmWave/THz signal.
Therefore, one can greatly reduce the pilot overhead by estimating the sparse channel parameters $\lbrace \theta_{i}^{l},\phi_{i}^{l},\tau_{i}^{l},\alpha_{i}^{l} \rbrace_{i=1}^{P}$ instead of the full-dimensional MIMO channel matrix $\mathbf{H}^{l}[k]$.
To this end, one can exploit the LSTM, a DL technique specialized for extracting temporally correlated features from the sequential data\footnote{Basically, the key ingredients of LSTM block are \textit{cell state} and three gates, viz., \textit{input gate}, \textit{forget gate}, and \textit{output gate}.
The cell state, serving as a memory to store information extracted from the past inputs, sequentially passes through the forget, input, and output gates.
Based on the input and the previous output, each gate determines how much information to be removed, written, and read in the cell state.}.
In our context, LSTM can dynamically adjust the channel parameter estimation based on the extracted temporally correlated feature.
For example, when the mobility of a mobile is low and thus the temporal correlation of channel parameters is high, past parameters highly affects the current parameters.
LSTM can effectively deal with this type of temporal dependency and finally makes a fast yet accurate channel parameter estimation with relatively small amount of pilot resources.
In a nutshell, LSTM-based channel estimator learns a complicated nonlinear mapping $g$ between the received downlink pilot signals ($\mathbf{y}^{1},\cdots,\mathbf{y}^{l}$) and the channel parameters ($\theta^l, \phi^l, \tau^l$, $\alpha^l$):
\begin{equation}
\{\hat{\theta}^l, \hat{\phi}^l, \hat{\tau}^l, \hat{\alpha}^l\} = g(\mathbf{y}^1, \cdots, \mathbf{y}^l;\Theta),
\end{equation}
where $\Theta$ is the set of weights and biases (see Fig.~\ref{fig_7}(a)).
To find out the parameters $\{ \hat{\theta}^l, \hat{\phi}^l, \hat{\tau}^l, \hat{\alpha}^l \}$, the channel estimate $\widehat{\mathbf{H}}^l[k] = \sum^{P}_{i=1} \hat{\alpha}^l_i e^{-j2 \pi k f_s \hat{\tau}_i^l} \mathbf{a}_\mathrm{R} (\hat{\theta}^l_i) \mathbf{a}_\mathrm{T} (\hat{\phi}^l_i)$ needs to be compared against the true channel matrix $\mathbf{H}^l[k]$ in the training.
To do so, we define the loss function $J(\Theta)$ as
\begin{align}
J(\Theta) = \frac{1}{L} \sum_{l=1}^{L} \sum_{k=1}^{K}  \Vert \mathbf{H}^l_k(\Theta) - \widehat{\mathbf{H}}^l_k(\Theta) \Vert_F^2,
\end{align}
where $L$ and $K$ are the number of coherence intervals and the number of subcarriers, respectively.

In order to evaluate the efficacy of the LSTM-based parametric channel estimation technique, we test the normalized MSE (NMSE) as the function of SNR\footnote{In our experiments, we assume the scenario that the user is moving along with the line trajectory so that the concept of the temporal correlation between geometric parameters can be reflected in the dataset.}.
As shown in Fig.~\ref{fig_7}(b), LSTM-based channel estimation technique outperforms the conventional channel estimation algorithms by a large margin since the trained DNN exploits the sparsity of the angular domain channel, but no such mechanism is used for the conventional LS and MMSE techniques.
For example, when SNR$=10\,$dB, the NMSE of the DL-based scheme is less than $10^{-2}$ while those of the LMMSE estimator and the LS estimator are $10^{-1}$ and $0.5$, respectively.
We also observe that the performance of DL-based scheme outperforms the CS-based approach (i.e., $5\,$dB gain at SNR$=15\,$dB), since the channel parameters can be readily estimated by using the temporal correlation between the past and current channels.

\subsection{CSI Feedback in mmWave/THz Massive MIMO System}
As mentioned, to fully enjoy the benefit of massive MIMO systems, an acquisition of an accurate downlink CSI at the BS is crucial.
To do so, UE needs to estimate the downlink channels using the pilot signal (e.g., CSI reference signal (CSI-RS) in 5G) of each antenna and then feeds them back in a form of implicit indices (e.g., PMI, RI, and CQI in 4G LTE and 5G NR) to the BS~\cite{TS38214}.
Since the required number of bits to convey these indices increases in proportion to the number of antennas, the CSI feedback overhead is a big concern for the massive MIMO systems.

\begin{figure}
	\centering
	\subfloat[]{\includegraphics[width=1\columnwidth]{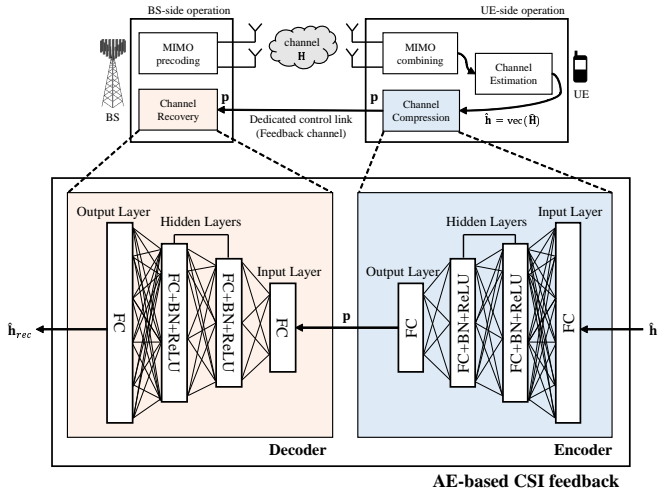}}
	\hfill
	\subfloat[]{\includegraphics[width=.85\columnwidth]{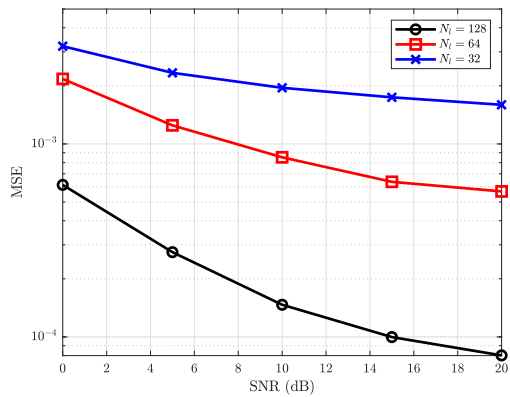}}
	\hfill
	\caption{(a) AE architecture for efficient CSI feedback. (b) MSE between the genie channel $\mathbf{H}$ and reconstructed channel estimate $\hat{\mathbf{H}}_{rec}$ as a function SNR. $N_l$ is the dimension of the latent vector $\mathbf{p}$.}
    \label{fig_sec3_c}
\end{figure}

Over the years, there have been many works to reduce the overhead of the CSI feedback by exploiting the spatio-temporal correlation of CSI~\cite{CSIfeedback1, CSIfeedback2, CSIfeedback3}.
For example, in~\cite{CSIfeedback1}, the correlated channel vectors are transformed into an uncorrelated sparse vector in some bases and then the CS technique is used to estimate the CSI from the small number of channel measurements.
Also, in recent years, DL-based CSI feedback schemes have been proposed~\cite{CSIfeedback5}, which are stimulated by the outstanding performance of DL in the correlation extraction.
DL-based feedback scheme can be derived from the autoencoder (AE) architecture (see Fig.~\ref{fig_sec3_c}(a)).
In the AE, \textit{encoder} is used to transform the raw channel matrices into a compressed codeword vector and \textit{decoder} reconstructs the original channel matrices from the codeword.

We consider the downlink MIMO-OFDM system where the BS is equipped with $N_{t}$ transmit antennas and the UE is equipped with $N_{r}$ receiver antennas.
Since each antenna port at the BS sends a CSI-RS for the channel measurement, $N_{t} \times N_{r}$ elements in the MIMO channel matrix $\mathbf{H}\in\mathbb{C}^{N_{r} \times N_{t}}$ should be fed back to the BS.
In the AE-based CSI feedback scheme, the encoder in the UE compresses the channel estimate $\hat{\mathbf{H}}$ into a low-dimensional feature vector and the decoder in the BS reconstructs the channel $\hat{\mathbf{H}}$ from the compressed vector.
As an input to the encoder, vectorized version of $\hat{\mathbf{H}}$ ($N_{t}N_{r} \times 1$-dimensional vector $\hat{\mathbf{h}}$) is used.
Then, we use multiple hidden layers to obtain the low-dimensional vector $\mathbf{p}$, which can be expressed as
\begin{align}
\mathbf{p} = f_{e}(\hat{\mathbf{h}}; \Theta_{e}),
\end {align}
where $f_{e}(\cdot)$ is the encoder operation and $\Theta_{e}$ is the training parameter set of the encoder.
In the decoder, to reconstruct $\hat{\mathbf{h}}$ using $\mathbf{p}$, we design multiple FCN layers with BN and ReLU activation (see Fig.~\ref{fig_sec3_c}(a)).
Finally, the reconstructed channel vector $\hat{\mathbf{h}}_{rec}$ can be expressed as
\begin{align}
\hat{\mathbf{h}}_{rec} = f_{d}(f_{e}(\hat{\mathbf{h}}; \Theta_{e});\Theta_{d}),
\end{align}
where $f_{d}(\cdot)$ is the decoder operation and $\Theta_{d}$ is the training parameter set of the decoder.
Considering that the training objective of the AE-based CSI feedback is to find out the channel closest to $\hat{\mathbf{h}}$, we use the MSE between the input channel and the reconstructed channel $\hat{\mathbf{h}}_{rec}$ as a loss function:
\begin{align}
    J(\Theta_{e},\Theta_{d})=\frac{1}{\sqrt{N_{t}N_{r}}}\left\Vert \hat{\mathbf{h}} - \hat{\mathbf{h}}_{rec} \right\Vert_{2}^{2}
\end{align}

In Fig.~\ref{fig_sec3_c}(b), we evaluate the MSE between the reconstructed channel $\hat{\mathbf{H}}_{rec}$ and the true channel $\mathbf{H}$ (i.e., $\left\Vert \hat{\mathbf{H}}_{rec} - \mathbf{H} \right\Vert_{2}^{2}$) of the AE-based CSI feedback scheme as a function of SNR.
Note that the genie channel $\mathbf{H}$ is obtained by the geometric MIMO channel model in~\eqref{eq:MIMO_channel}.
As shown in Fig.~\ref{fig_sec3_c}(b), we observe that the reconstruction quality of AE-based scheme improves gradually with the dimension $N_l$ of the latent vector $\mathbf{p}$.
This is mainly because the latent vector having a large dimension can capture the detailed channel features (e.g., geometric channel parameters of NLoS paths) as well as the core features (e.g., channel parameters of the dominating LoS path).
For example, we observe that the AE-based CSI feedback technique achieves MSE$=10^{-4}$ at SNR = 15 dB when $N_{l} = 128$, which is by no means possible in other cases even in the higher SNR regime.

\subsection{Meta Learning-based GAN for Real-time Training}
In Section III.A, we discussed that the GAN technique can be used to collect realistic wireless channel samples.
In reality, however, GAN is also a data-driven DL technique so that it requires considerable training samples and hence the practical benefit of this approach might be washed away when the training data is insufficient~\cite{jhkim_GAN}.
To overcome the shortcoming, one can consider the \textit{meta learning}, a technique to train a model using a variety of tasks and then solve a new task using only a small number of training samples~\cite{finn2017model}.
In short, the meta learning is a special training technique to obtain the initialization parameters of DNN using which one can easily and quickly learn the desired function with small training samples.

\begin{algorithm}[t!]
    \caption{Meta Learning-based GAN Training Process}
  \begin{algorithmic}[1]
    \INPUT Wireless channel dataset $\{D_i\}^{M+1}_{i=1}$, learning rates $\alpha$, $\beta$, and $\gamma$
    \STATE randomly initialize GAN parameters $\theta$
    \WHILE{meta learning}
        \FOR{\textbf{all} i}
        \STATE Sample batch data $\mathbf{x}^{(i)}$ from $D_i$
        \STATE Evaluate $\nabla_\theta \mathcal{L}_{D_i}$ using $\mathbf{x}^{(i)}$ and GAN loss $\mathcal{L}_{D_i}$
        \STATE Compute adapted parameters with gradient descent: $\psi_{D_i} = \theta - \alpha \nabla_\theta \mathcal{L}_{D_i}(\theta)$
        \STATE Sample batch data for meta-update $\mathbf{x'^{(i)}}$ from $D_i$
        \ENDFOR
    \STATE Update $\theta = \theta - \beta \nabla_\theta \sum_{i=1}^{M} \mathcal{L}_{D_i}(\psi_{D_i})$ using $\mathbf{x'^{(i)}}$ and GAN loss $\mathcal{L}_{D_i}$
    \ENDWHILE
    
    \WHILE{parameter update}
    \STATE Sample batch data $\mathbf{x}^{(M+1)}$ from $D_{M+1}$
        \STATE Evaluate $\nabla_\theta \mathcal{L}_{D_{M+1}}$ using $\mathbf{x}^{(M+1)}$ and GAN loss $\mathcal{L}_{D_{M+1}}$
        \STATE Compute adapted parameters with gradient descent: $\theta = \theta - \gamma \nabla_\theta \mathcal{L}_{D_{M+1}}(\theta)$
    \ENDWHILE
  \end{algorithmic}
  \label{algorithm}
\end{algorithm}

\begin{figure*}
	\centering
	\subfloat[]{\includegraphics[width=1.18\columnwidth]{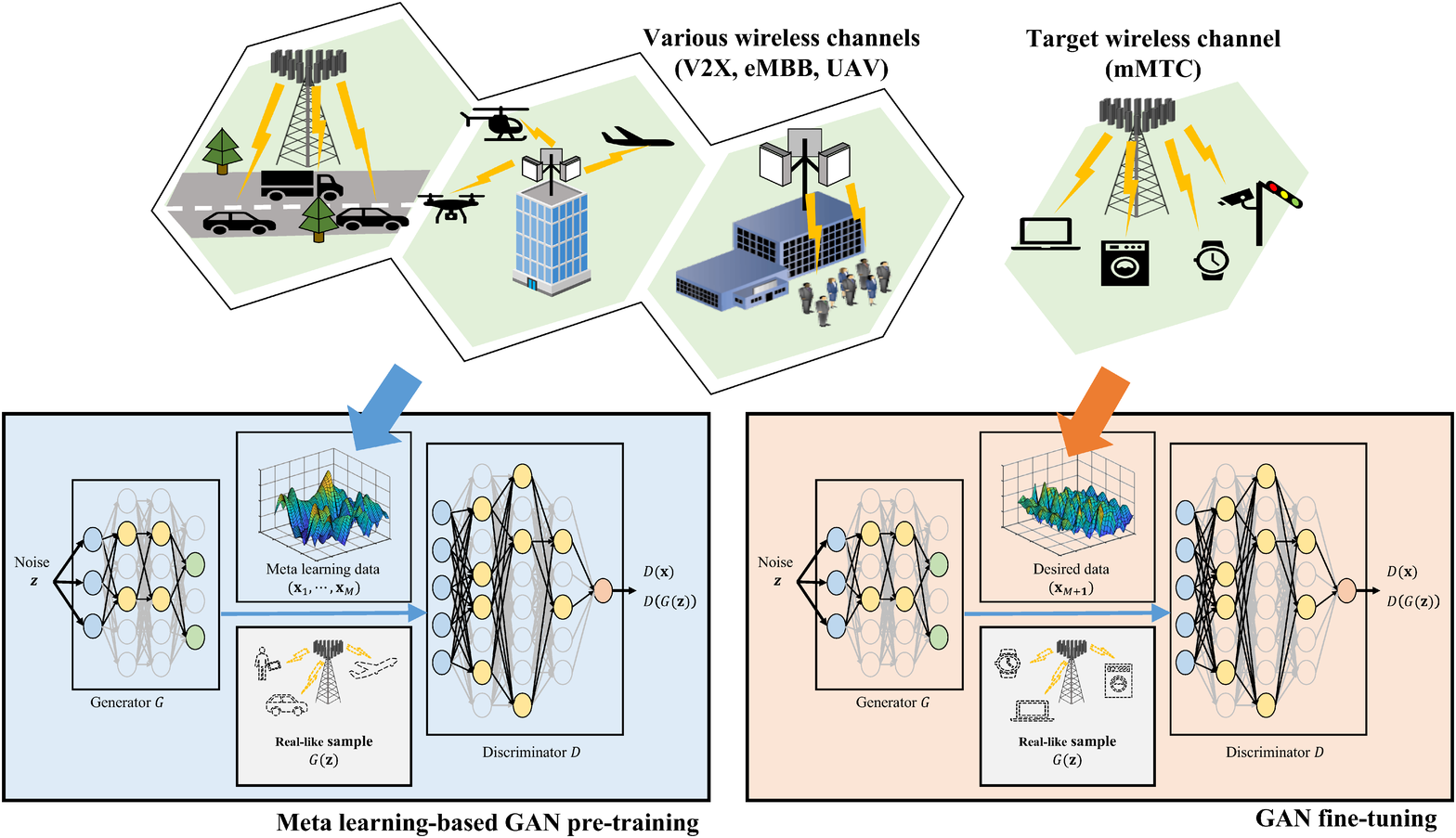}}
	\hfill
	\subfloat[]{\includegraphics[width=.82\columnwidth]{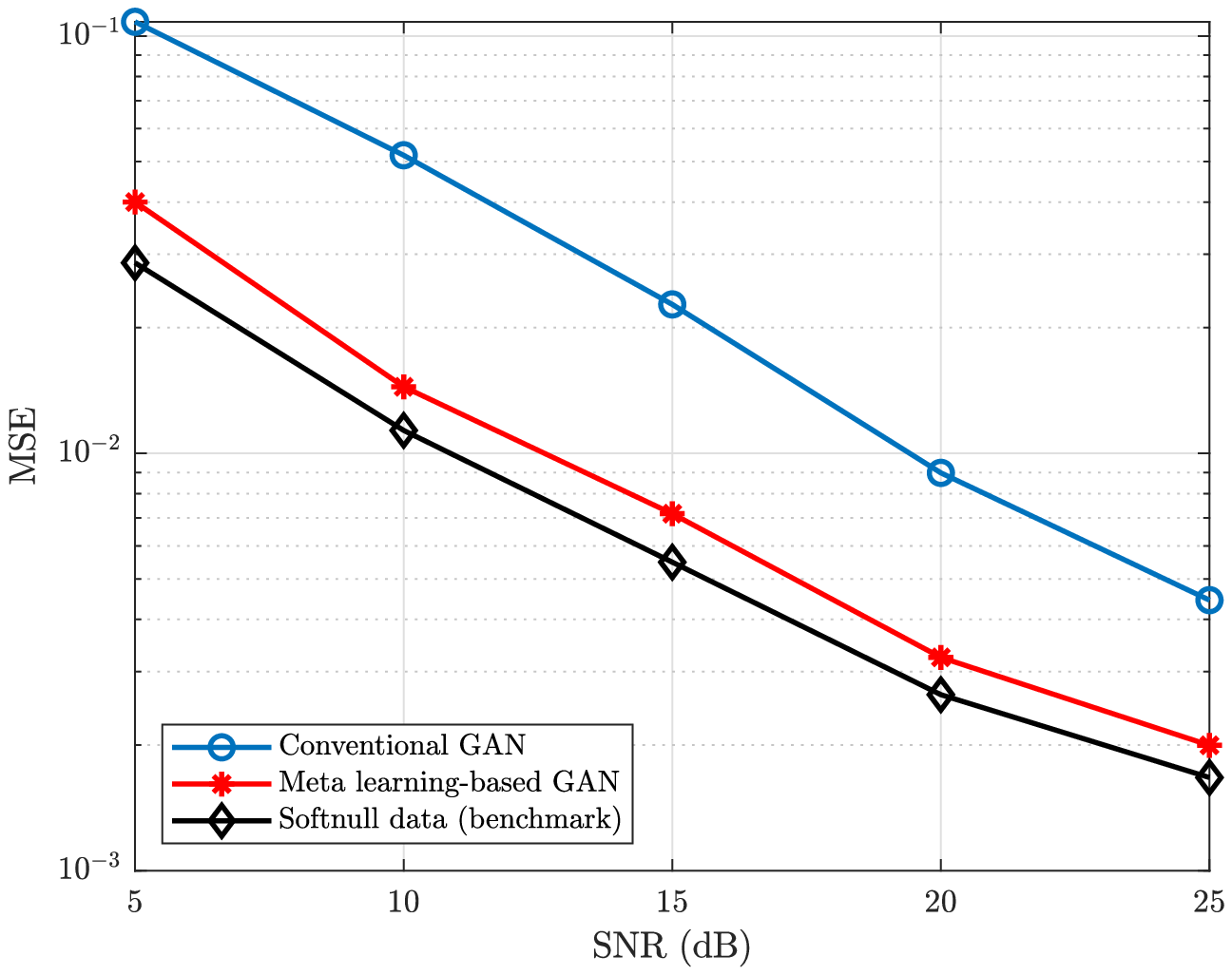}}
	\hfill
	\caption{(a) Architecture of the meta learning-based GAN pretraining and fine-tuning framework (b) MSE performances of the DL-based channel estimator using three distinct datasets: real measured dataset, generated dataset from conventional GAN and GAN trained via meta learning}
	\label{fig_9}
\end{figure*}

To be specific, an overall procedure of the meta learning-based GAN training is as follows.
First, we perform the meta learning to obtain the initialization parameters.
We then update the network parameters of GAN to perform the fine-tuning of DNN such that trained DNN generates channel samples for the desired wireless environments.
In the meta learning phase, we extract the common features of multiple channel datasets, say $M$ datasets $\left\{ D_{1}, \cdots, D_{M}\right\}$, and then use them to obtain the network initialization parameters $\theta$ (see Fig.~\ref{fig_9}(a)):
\begin{align}
    \psi_{D_i, t} &= \theta_{t-1} - \alpha \nabla_\theta \mathcal{L}_{D_i}(\theta_{t-1}),\\
    \theta_{t} &= \theta_{t-1} - \beta \nabla_\theta \sum_{i=1}^{M} \mathcal{L}_{D_i}(\psi_{D_i,t}),
\end{align}
where $\theta_t$ and $\theta_{t-1}$ are the parameters of GAN in $t$-th step and $(t-1)$-th step, respectively.
Also, $\psi_{D_i, t}$ is the parameter associated with dataset $D_i$ in $t$-th step, $\mathcal{L}_{D_i}$ is the loss function of GAN for $i$-th dataset $D_i$, and $\alpha$ and $\beta$ are the step sizes for the parameter update (see Algorithm~\ref{algorithm}).
Next, in the parameter update phase, we use $\theta$ as the initialization parameters and then train DNN to generate the samples close to the desired channel dataset, say $D_{M+1}$.
In a nutshell, all that needed is to learn the distinct features (of $D_{M+1}$) unextracted from the meta learning.

In Fig.~\ref{fig_9}(b), we test the MSE performances of the DL-based channel estimator trained by the three different datasets: real-measured dataset~\cite{softnull}, dataset generated from the vanilla GAN and GAN trained by meta learning\footnote{Specifically, we have used 10,000 samples for the benchmark training, 4,000 samples for meta learning, and 800 samples for fine-tuning and the training of the conventional CGAN.}.
We observe that the channel estimation performance of the meta learning-based GAN is slightly worse than that using the real samples (e.g., 2 dB loss at MSE=$10^{-2}$).
Whereas, the performance of the conventional GAN-based approach is much worse (i.e., more than 6 dB loss at MSE=$10^{-2}$) since the number of training samples is not large enough to ensure the convergence of GAN and there is no mechanism to exploit the common features of diverse channel conditions.

\section{Conclusion and Discussion}
In this article, we discussed the DL-based channel estimation with emphasis on the design issues related to DL model selection, training set acquisition, and DNN architecture design.
As the automated services and applications using machines, vehicles, and sensors proliferate, we expect that DL will be more popular channel estimation paradigm in 6G era.
To deal with various frequency bands (sub-6GHz/mmWave/THz), wireless resources (massive MIMO antennas, intelligent reflecting surface, relays), and geographical environment, we need to go beyond the state-of-the-art DL technique and consider more aggressive and advanced DL techniques.
For example, when we try to train a DL model to estimate the desired wireless channel, transfer learning, an approach to use the pre-trained model for a similar task, can be employed.
By recycling most of parameters in the pre-trained model and then training only a small part of parameters, new model can learn the distinct feature of the desired channel while reusing common features of the pre-trained channel environments.
Another approach worth investigation is the federated/split/distributed learning, a technique to learn the desired task by the cooperation of multiple decentralized devices or servers.
Our hope is that this article will serve as a useful reference for the communication researchers who want to apply the DL technique in their wireless channel estimation application.

\section*{Acknowledgment}
% This research was supported by the MSIT(Ministry of Science and ICT), Korea, under the ITRC(Information Technology Research Center) support program(IITP-2022-2017-0-01637) supervised by the IITP(Institute for Information \& Communications Technology Planning \& Evaluation) and Electronics and Telecommunications Research Institute (ETRI) grant funded by the Korean government [16ZI1100, Wireless Transmission Technology in Multi-point to Multi-point Communications].
This work was partly supported by Institute of Information \& communications Technology Planning \& Evaluation (IITP) grant funded by the Korea government (MSIT) (No.2021-0-00972, Development of Intelligent Wireless Access Technologies) and the ITRC (Information Technology Research Center) support program (IITP-2022-2017-0-01637) supervised by the IITP.

\ifCLASSOPTIONcaptionsoff
  \newpage
\fi

\bibliographystyle{IEEEtran}

\vspace{-1cm}

\end{document}